\documentclass[12pt]{article}
\usepackage[utf8]{inputenc}
\usepackage{amsmath}
\usepackage{amssymb}
\usepackage{graphicx}

\usepackage{comment}
\newcommand{\ntime}{n}

\makeatletter

\DeclareTextSymbolDefault{\textquotedbl}{T1}

\numberwithin{equation}{section}

\@ifundefined{date}{}{\date{}}

\usepackage{braket}
\usepackage{bm}

\usepackage{epsfig}
\usepackage{pstricks}
\usepackage{color}
\usepackage{tikz}

\usepackage{hyperref}

\newcommand {\nn} {\nonumber}
\newcommand{\bbR}{{\mathbb R}}
\newcommand{\bbC}{{\mathbb C}}

\@ifundefined{showcaptionsetup}{}{%
\PassOptionsToPackage{caption=false}{subfig}}
\usepackage{subfig}
\makeatother

\begin{document}
\begin{titlepage}
\renewcommand{\thefootnote}{\fnsymbol{footnote}}

\begin{flushright} 
KEK-TH-2640
\end{flushright} 

\vspace{2cm}

\begin{center}
    {\bf \large Monte Carlo studies of
    quantum cosmology\\
   by the generalized Lefschetz thimble method}
\end{center}

\vspace{0.5cm}


\begin{center}
        Chien-Yu C{\sc hou}$^{1,2)}$\footnote
        { E-mail address : ccy@post.kek.jp} and
        Jun N{\sc ishimura}$^{1,2)}$\footnote
        { E-mail address : jnishi@post.kek.jp}


\vspace{1cm}

$^{1)}$\textit{KEK Theory Center,
Institute of Particle and Nuclear Studies,}\\
{\it High Energy Accelerator Research Organization,\\
1-1 Oho, Tsukuba, Ibaraki 305-0801, Japan} 


~

$^{2)}$\textit{Graduate Institute for Advanced Studies, SOKENDAI,\\
1-1 Oho, Tsukuba, Ibaraki 305-0801, Japan}

\end{center}

\vspace{1cm}

\begin{abstract}
 \noindent
 Quantum cosmology
 aims at elucidating
 the beginning of our Universe.
 Back in early 80's,
 Vilenkin and Hartle-Hawking
 put forward
 the ``tunneling from nothing'' and ``no boundary'' proposals.
 Recently
 there has been renewed interest in this subject
 from the viewpoint of defining the oscillating path integral
 for Lorentzian quantum gravity using the Picard-Lefschetz theory.
 Aiming at going beyond the mini-superspace
 and saddle-point approximations,
 we perform
  Monte Carlo calculations
 using the generalized Lefschetz thimble method
 to overcome the sign problem.
 In particular, we confirm that
 either the Vilenkin or the Hartle-Hawking saddle point
 becomes relevant
 if one uses the Robin boundary condition
 depending on its parameter.
 We also clarify some fundamental issues in quantum cosmology,
 such as
 an issue related to the integration domain of the lapse function.
\end{abstract}
\vfill
\end{titlepage}
\vfil\eject


\setcounter{footnote}{0}

\section{Introduction}

Quantum cosmology is a fascinating subject, which aims at elucidating
the beginning of our universe.
In early 80's, two important ideas towards this goal were put forward.
One is known as the ``tunneling from nothing'' proposal by
Vilenkin \cite{Vilenkin:1982de,Vilenkin:1984wp,Vilenkin:1994rn},
which states that our universe is
born through quantum tunneling
from an initial quantum state with no classical geometry.
The other is known as the ``no boundary'' proposal
by Hartle and Hawking \cite{Hartle:1983ai},
which states that the wave function of our universe
should be given by path integral over Euclidean geometries
without a boundary corresponding to the initial time.
Vilenkin's proposal can also be stated in
the path integral formalism \cite{Vilenkin:1984wp},
in which one integrates over Lorentzian geometries
unlike Hartle-Hawking's proposal.
%
Despite their conceptual similarities,
the two proposals lead to quite different consequences.
For instance, the wave functions obtained by the Vilenkin and Hartle-Hawking
proposals
behave as
$e^{-12 \pi^2 / \ell_{\rm P}^2 \Lambda}$ and $e^{12 \pi^2 / \ell_{\rm P}^2 \Lambda}$,
respectively, where
$\Lambda$ is the cosmological constant and
$\ell_{\rm P}$ is the Planck length. 
This makes a big difference when
the cosmological constant is replaced by the potential of a dynamical field
as in the inflation models.
See Refs.~\cite{Lehners:2023yrj,Maldacena:2024uhs}
for recent reviews
on these proposals
and related issues.

The relationship between the two proposals
can be
most clearly seen \cite{Halliwell:1988ik}
in the mini-superspace model, which
is obtained from the Einstein-Hilbert action by assuming
that the geometry to be integrated over is homogeneous and isotropic.
After integrating out the scale factor
while fixing the time-reparametrization
invariance by requiring the lapse function to be time-independent,
one is left with the integration over the lapse $N$.
The saddle points on the complex $N$ plane that correspond to
Vilenkin and Hartle-Hawking were identified,
and they actually correspond to different orientations of the Wick rotation;
namely the usual one $t=-i\tau$ for the Hartle-Hawking saddle
and the opposite one $t=i\tau$ for the Vilenkin saddle.
Possible choice of the
steepest descent contours
that go through
either the Vilenkin or the Hartle-Hawking saddle was also discussed,
but there was no principle to choose
a unique contour at that time.

%

Recently there has been
major progress 
in defining the path integral for quantum gravity
based on the 
Picard-Lefschetz theory, which
provides a general prescription for rendering an oscillating integral
into a sum over convergent integrals on the Lefschetz thimbles associated
with relevant saddle points.
Using this theory,
the real-time path integral
can be made totally well defined without
making
a Wick rotation into the imaginary time formalism.
In particular, this
has a big impact on the formulation of quantum gravity
since the Euclidean quantum gravity is known to be pathological
due to the unboundedness of the Einstein action
as is manifested by the conformal mode instability \cite{Gibbons:1978ac}.
Such pathology is absent in Lorentzian quantum gravity
since the unboundedness simply implies increasingly violent phase rotations
of the integrand, which can be dealt with by the Picard-Lefschetz theory.
Therefore, a natural definition
of the path integral for quantum
gravity can be
given by integrating over the Lorentzian geometry
as advocated in Ref.~\cite{Feldbrugge:2017kzv}.
In particular, this fixes the ambiguity\footnote{In this paper,
we are primarily concerned with the propagator (\emph{i.e.}, Green’s function),
which is relevant for dynamical (Lorentzian) processes.
However, in the context of using the path integral for the purpose of constructing
the wave function as a solution to the Wheeler-DeWitt equation,
there is no reason to prefer a Lorentzian contour. Hence, the ambiguity still remains in
that context \cite{DiazDorronsoro:2017hti,DiazDorronsoro:2018wro,Halliwell:2018ejl}.}
in
choosing the
integration contour in quantum cosmology
completely.\footnote{Recently
  there are interesting observations related to this issue
  from the viewpoint of the ${\rm dS}_3/{\rm CFT}_{2}$
  correspondence \cite{Chen:2023prz,Chen:2023sry,Chen:2024vpa,Chen:2024qmn}.} 

In fact, 
the steepest descent contour
specified
in this way
passes through the Vilenkin saddle when the Dirichlet
boundary condition is imposed at the initial time.
However, 
it has been argued that
the Vilenkin saddle suffers from the instability problem
if one incorporates
the tensor modes and matter
fields because of the ``wrong'' Wick rotation mentioned
above \cite{Feldbrugge:2017fcc,Feldbrugge:2017mbc}.
(See also Refs.\cite{Vilenkin:2018dch,Vilenkin:2018oja,Feldbrugge:2018gin,Matsui:2022lfj,Matsui:2023hei,Matsui:2024bfn}.)
Based on these observations,
it has been proposed \cite{DiTucci:2019dji,DiTucci:2019bui}
to use the Robin boundary condition\footnote{Historically,
  the Robin boundary condition was introduced in quantum cosmology
  earlier \cite{Vilenkin:2018dch,Vilenkin:2018oja}
  in order to solve the instability problem of the Vilenkin saddle point.}
at the initial time,
which interpolates the Dirichlet and Neumann conditions
while making
the scale factor vanish on the average at the initial time.
In that case,
the relevant saddle point has been shown to
switch from Vilenkin to Hartle-Hawking as the interpolating parameter
of the Robin boundary condition is changed.
Thus the no-boundary proposal with the Hartle-Hawking saddle point
was realized, for the first time, within Lorentzian quantum gravity.


In this paper,
we perform
Monte Carlo calculations in
quantum cosmology.
In order to overcome the sign problem that occurs
for an oscillating integral,
we apply
the generalized Lefschetz thimble method (GTM) \cite{Alexandru:2015sua},
which ameliorates
the sign problem by
deforming the integration contour similarly to the Picard-Lefschetz
theory.\footnote{The idea to perform Monte Carlo simulations
  on the Lefschetz thimble was put forward in
  Refs.~\cite{Witten:2010cx,Cristoforetti:2012su,Cristoforetti:2013wha,Fujii:2013sra}.}
A first step in that direction has been taken
recently\footnote{This paper addressed
the issue of the integration domain of the scale factor squared $q(t)=a^2(t)$,
which is taken to be the whole real axis
in analytic calculations such as the one in Ref.~\cite{Halliwell:1988ik}.
For other numerical approaches to quantum gravity,
see Refs.~\cite{Loll:2019rdj,Jia:2021xeh}, for instance.}
using a simple Metropolis algorithm on the real axis \cite{Jia:2022nda}.
Here we use a far more
efficient Hybrid Monte Carlo
algorithm on the deformed contour \cite{Fukuma:2019uot},
which is crucial, in particular, in investigating
interesting behaviors
that appear 
in the case of Robin boundary condition.
We also use the
preconditioned
flow equation \cite{Nishimura:2024bou}
to overcome some technical problem that occurs when
the number of time steps is increased.
This technique has been used recently in establishing
a new picture of quantum tunneling
based on
the real-time path integral
formalism \cite{Nishimura:2023dky}.


The primary aim
of this work is to establish a
calculation method
for quantum cosmology
that enables us to perform fully nonperturbative calculations
that goes beyond the saddle-point approximation.
Furthermore, while we restrict ourselves to the mini-superspace model in this work,
the established
method is flexible enough to
add matter fields and/or to enlarge the mini-superspace
by expanding fields in spherical harmonics and keeping a finite number of them,
which is important
in investigating
the possible
instability problem with the Vilenkin saddle mentioned above.

In particular, our numerical calculations
confirm that the Vilenkin saddles become relevant
in the case of Dirichlet boundary condition.
The emergence of the real time is described by the
Stokes phenomenon of the saddle points,
which occurs as we increase the scale factor at the final time.
In the case of Robin boundary condition,
the relevant saddle point switches
from Vilenkin to Hartle-Hawking as the interpolating parameter
of the Robin boundary condition is changed.
These results are obtained
in a regime in which perturbation theory around
saddle points breaks down as we show by explicit calculations.

We also clarify some fundamental issues in quantum cosmology,
which have not been fully discussed in the literature.
In particular, we discuss some issues concerning
the integration domain of the lapse function.
In fact, the whole real axis has to be chosen
in order to obtain solutions to the Wheeler-DeWitt equation,
whereas
the positive real axis has to be chosen in order to calculate the Green functions
of the Wheeler-DeWitt
operator \cite{Teitelboim:1981ua,Teitelboim:1983fh,Teitelboim:1983fk}.
(See Ref.~\cite{Banihashemi:2024aal} for a recent discussion.)
In the case of Robin boundary condition, however,
it turns out that there is a problem in
choosing the positive real axis.
This is due to the fact that
the origin $N=0$ is mapped to a different point
by the contour deformation
unlike in the case of Dirichlet boundary condition.
Because of this,
one obtains an unwanted contribution from the arc
that starts at the origin $N=0$ when
one deforms the integration contour using Cauchy's theorem.
We determine the parameter regime in which this contribution
dominates over the
contribution from the relevant saddle.

The rest of this paper is organized as follows.
In Section \ref{sec:mini-superspace}, we briefly review
quantum cosmology based on the mini-superspace
model
with the Dirichlet and Robin boundary conditions.
In Section \ref{ap:arc},
we discuss the issue of the arc contribution that appears in the case of Robin boundary condition.
In Section \ref{sec:GTM-qc}, we explain how we deal with this model by
Monte Carlo calculations using the GTM.
In Section \ref{sec:results}, we present our results obtained
by the Monte Carlo calculations.
In Section \ref{sec:geo},
we discuss how one can read off
the real geometry
from the complex geometry obtained at the saddle point.
Section \ref{sec:summary} is devoted to a summary and discussions.
In Appendix \ref{sec:GTM}, we explain the details of our calculation method.
In Appendix \ref{sec:eff-action-epsilon}, we
derive the effective action for the lapse function
in the theory with discrete time, which is used in perturbative calculations.

\section{Brief review of quantum cosmology}
\label{sec:mini-superspace}

In this section, we briefly review quantum cosmology
based on the mini-superspace model, in which
the space-time is assumed to be homogeneous and isotropic.
Here we also assume that it has positive curvature.
Then the metric can be parametrized as
\begin{eqnarray}
ds^2 = a^2(\eta) (-N(\eta)^2 d\eta^2 + d \Omega_3{}^2 ) \ ,
\label{metric-a-eta}
\end{eqnarray}
where we have defined the conformal time $\eta$,
the scale factor $a(\eta)$,
the lapse function $N(\eta)>0$
and the metric $d \Omega_3{}^2$ on a unit 3-sphere.
Plugging this metric into the Einstein-Hilbert action
with the Gibbons–Hawking–York boundary term, we
obtain\footnote{Throughout this paper,
  we set $\hbar = c = 8 \pi G_{\rm N} = 1$, which corresponds to
  setting the Planck length to $l_{\rm P}=\sqrt{\frac{\hbar G_{\rm N}}{c^3}}
    = \frac{1}{\sqrt{8\pi}}$.}
\begin{eqnarray*}
S_{\rm EH}[a,N]= 6 \pi^2 \int d\eta
\left\{ - \frac{1}{N^2}a \dot{a}^2
 + N (a^2 - H^2 a^4) \right\} \ ,
\end{eqnarray*}
where
$H^2$ represents the cosmological constant $\Lambda \equiv 3 H^2$
using the notation in Refs.~\cite{DiTucci:2019dji,DiTucci:2019bui}.

In order to solve this model analytically, we make a change of
variables \cite{Halliwell:1988ik}
\begin{align}
q(t) &= a^2 (\eta)  \ , \nn \\
dt &= a^{2} (\eta) \, d\eta \ ,
  \label{q-a-relation}
\end{align}
which brings the action into a quadratic form
\begin{eqnarray}
S_{\rm EH}[q,N]= 6 \pi^2 \int dt\left\{-\frac{1}{4N} \dot{q}^2+ N (1-H^2 q) \right\}
\label{action-EH}
\end{eqnarray}
with respect to $q$, where we have introduced
$\dot{q}=\frac{dq}{dt}$ and
$\ddot{q}=\frac{d^2 q}{dt^2}$.
In terms of the new variables, the metric \eqref{metric-a-eta} reads
\begin{eqnarray}
 ds^2 = -\frac{N(t)^2}{q(t)} dt^2 + q(t) \, d \Omega_3{}^2 \ .
\label{metric-q-t}
\end{eqnarray}




Following Ref.~\cite{Halliwell:1988ik},
we quantize this theory
by the path integral in terms of
the canonical variables
including
the conjugate momenta $p(t)$ and $\pi(t)$ for $q(t)$ and $N(t)$, respectively.
After fixing the time-reparametrization invariance
by using the
constant-lapse gauge $\dot{N}=0$
and integrating out
the associated Batalin-Fradkin-Vilkovisky ghosts as well as $\pi(t)$,
we arrive at the partition function
\begin{alignat}{1}
  Z&= \int dN \, {\cal D}q \, {\cal D}p\,e^{iS} \ , 
  \label{def-Z-mini-sup-0}
  \\
S[q,p,N]&= \int dt
\left[ \dot{q}p+ N \left\{ {1\over 6\pi^2}p^2+6\pi^2(1-H^2q) \right\} \right] \ .
           \label{eq:S-q-p-N-0}
\end{alignat}

Here we note that
$H^2$ that appears in \eqref{eq:S-q-p-N-0}
can be pulled out of the action as
\begin{alignat}{1}
  \tilde{Z} &= \int d\tilde{N} \, {\cal D}\tilde{q} \, {\cal D}p\,e^{i\tilde{S}/H^2}  \ , 
  \label{def-Z-mini-sup}
  \\
\tilde{S}[\tilde{q},p,\tilde{N}]&= \int dt
\left[ \dot{\tilde{q}}p+
  \tilde{N} \left\{ {1\over 6\pi^2}p^2+6\pi^2(1-\tilde{q}) \right\} \right] \ ,
           \label{eq:S-q-p-N}
\end{alignat}
by rescaling $\tilde{q}(t) =H^2 q(t)$, $\tilde{N} = H^2 N$.
Thus we find that $H^2$ plays the role of $\hbar$ in quantum mechanics,
which controls the strength of quantum effects.
In this section, we consider the $H^2\rightarrow0$ limit with fixed
$\tilde{q}(t) =H^2 q(t)$, $\tilde{N} = H^2 N$,
which enables us to solve the theory \eqref{def-Z-mini-sup}
by the saddle-point analysis.
Remembering that
we deal with the rescaled variables $\tilde{q}$ and $\tilde{N}$ from now on,
we omit the tildes to simplify the notation.




\subsection{Dirichlet boundary condition}
\label{sec:DD-bc}

Integrating out $p(t)$ in \eqref{def-Z-mini-sup},
we obtain the effective action for $q(t)$ and $N$ as
\begin{alignat}{1}
    S_{\rm eff}[q,N]&=
  6\pi^2
        \int dt\,\left\{ -{1\over4N}\dot{q}^2+N(1-q)\right\} \ .
\label{action-qn}
\end{alignat}
Taking the variation of \eqref{action-qn} with respect to $q(t)$, we get
\begin{alignat}{1}
   \delta   S_{\rm eff} &=
         6 \pi^2
        \left[
   \int_0^1 dt \left( \frac{1}{2N} \ddot{q} - N \right) \delta q(t)
   - \frac{1}{2N} \Big\{ \dot{q}(1) \delta q(1)  - \dot{q}(0) \delta q(0) \Big\}
   \right] \ ,
\label{delta-S-EH}
\end{alignat}
from which one obtains the classical equation of motion for $q(t)$ as
\begin{alignat}{1}
  \ddot{q}=2N^2
\label{qEOM}
\end{alignat}
and the requirement for the boundary conditions
\begin{alignat}{1}
  \dot{q}(1) \delta q(1) - \dot{q}(0) \delta q(0) = 0
  \label{eq:require-bc}
   \ .
\end{alignat}

Here we impose
the Dirichlet boundary conditions
\begin{alignat}{1}
 q(0) =q_{\rm i} \ , \qquad
 q(1) =q_{\rm f}
\label{Dbc}
\end{alignat}
on both ends, which satisfy \eqref{eq:require-bc}.
Solving the classical equation of motion for $q(t)$ in \eqref{qEOM}
with the boundary conditions \eqref{Dbc}, we obtain
\begin{alignat}{1}
  \bar{q}(t) &= N^2 t^2 + ( - N^2 + q_{\rm f}
  -q_{\rm i}
  )\, t 
   + q_{\rm i} \ .
  \label{Dqcl}
\end{alignat}

The effective action for $N$ after integrating out $q(t)$
can be obtained in the $H^2\rightarrow0$ limit
by just substituting the classical solution
$q(t) = \bar{q}(t)$ in the action \eqref{action-qn}
as\footnote{Note that
the $N$-dependent factors
  arising from integrating out $p(t)$ and $q(t)$
  need not to be included in the effective action
  \eqref{action-qn} and \eqref{eq:eff-S-DD}
in the $H^2\rightarrow0$ limit considered here
although it should be taken into account
in the full quantum calculations as we discuss
in Section \ref{sec:GTM-qc}.} 
\begin{alignat}{1}
  S_{\rm eff}(N)
  = \pi^2\left\{ {N^3\over2} - 3N(q_{\rm f}+q_{\rm i}-2)
  - {3\over2N}(q_{\rm f}-q_{\rm i})^2\right\} \ .
  \label{eq:eff-S-DD}
\end{alignat}
Taking the derivative of \eqref{eq:eff-S-DD} with respect to $N$,
we obtain the saddle-point equation
\begin{alignat}{1}
N^4 - 2 (q_{\rm f}+q_{\rm i}-2)N^2 + (q_{\rm f}-q_{\rm i})^2  = 0 \ ,
  \label{eq:SPE-N-DD}
\end{alignat}
which can be factorized as
\begin{alignat}{1}
  \Big\{  N^2 -2 \sqrt{q_{\rm i} -1 }N  - (q_{\rm f}-q_{\rm i})  \Big\}
  \Big\{  N^2 +2 \sqrt{q_{\rm i} -1 }N  - (q_{\rm f}-q_{\rm i})  \Big\} = 0 \ .
  \label{eq:SPE-N-DD-2}
\end{alignat}
Thus we obtain four solutions\footnote{Here we define
  $\sqrt{q_{\rm f}-1}\equiv i \sqrt{1-q_{\rm f}}$ for $q_{\rm f}<1$
  and similarly for $\sqrt{q_{\rm i}-1}$.}
\begin{alignat}{1}
\bar{N}&=  c_1\sqrt{q_{\rm f}-1} +  c_2 \sqrt{q_{\rm i}-1}  \ ,
\label{DNcl}
\end{alignat}
where $c_1,\,c_2\in\{1,-1\}$.
Let us label them as $\bar{N}_1,\,\bar{N}_2,\,\bar{N}_3,\,\bar{N}_4$
corresponding to $(c_1,c_2) = (1,1)$, $(-1,1)$, $(-1,-1)$, $(1,-1)$,
respectively.

\begin{figure}[t]
   \centering
   \includegraphics[width=0.49\textwidth]{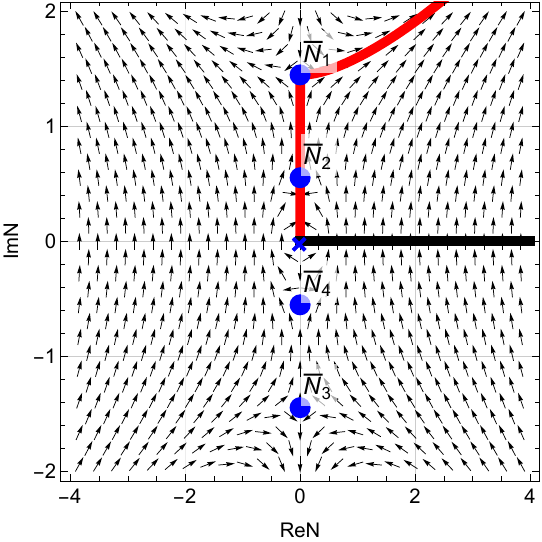}
   \includegraphics[width=0.49\textwidth]{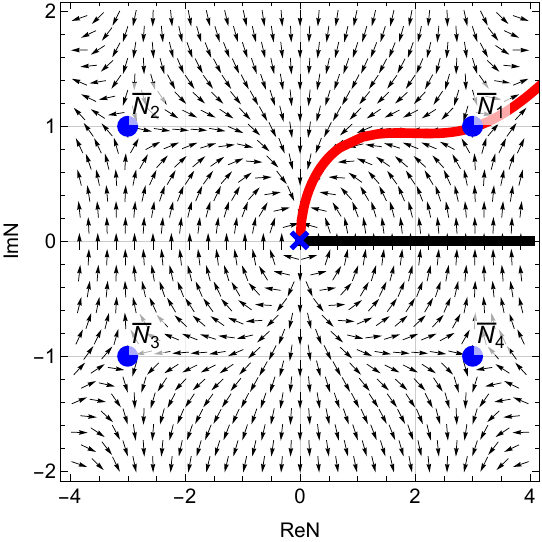}
   \caption{The flow diagram is shown in the case of
Dirichlet boundary conditions $q_{\rm i}=0$ at $t=0$
and $q_{\rm f}=0.8$ (Left), $10$ (Right) at $t=1$, respectively.
Blue circles represent the saddle points
$\bar{N}_1,\,\bar{N}_2,\,\bar{N}_3,\,\bar{N}_4$,
which correspond to $(c_1,c_2) = (1,1)$, $(-1,1)$, $(-1,-1)$, $(1,-1)$ 
in (\ref{DNcl}), respectively.
The blue crosses represent the singularities of the effective action.
The black line represents the original integration contour, while
the red line represents the integration contour obtained
by the Picard-Lefschetz theory.}
   \label{flowchart:D}
\end{figure}

In what follows, we set $q_{\rm i}=0$ in order to
realize the ``no boundary'' proposal.
Then the saddle points becomes 
\begin{alignat}{1}
\bar{N}&=  c_1\sqrt{q_{\rm f}-1} +  i  c_2  \ .
\label{DNcl-2}
\end{alignat}
Using $\bar{N}^2 -2i c_2 \bar{N} - q_{\rm f}=0$
in \eqref{Dqcl},
the classical solution $\bar{q}(t)$ can be rewritten as
\begin{alignat}{1}
  \bar{q}(t) &= \bar{N}^2t^2-2 i c_2 \bar{N}t \ .
 \label{cl:Dqn}
\end{alignat}

The integration contour for the lapse function $N$ can be obtained
by the Picard-Lefschetz theory, which amounts to deforming
the original integration contour $\bbR_{+}$
in the complex $N$ plane 
using the flow equation\footnote{This is nothing but 
the anti-holomorphic gradient flow equation \eqref{floweq}
applied to a single variable $N$ with the action $S(N) = -i \, S_{\rm eff}(N)$.}
\begin{alignat}{1}
\frac{\partial N(\tau)}{\partial \tau}
=i \, \overline{\partial S_{\rm eff}( N(\tau))\over\partial N} \ ,
\label{floweq-for-N}
\end{alignat}
where
$N(\tau)$ represents the lapse function at the flow time $\tau$
and $S_{\rm eff}(N)$ represents the effective action \eqref{eq:eff-S-DD}.
The initial condition $N(0)$ of the flow equation
is given by some point on the original integration contour $\bbR_{+}$.
The deformed integration contour obtained 
in the $\tau \rightarrow \infty$ limit passes through 
some saddle points, which are considered relevant.\footnote{See
  Ref.~\cite{Honda:2024aro} for a careful analysis in this regard
  from the viewpoint of the resurgence theory.}

In Fig.~\ref{flowchart:D}, we plot the right-hand side 
of \eqref{floweq-for-N} as
small arrows
in the complex $N$ plane.
The black line and the red line represent
the original integration contour and the deformed integration contour,
respectively.
For
$q_{\rm f} < 1$ (Left),
$\bar{N}_1$ and $\bar{N}_2$ are relevant with the latter being the dominant one,
whereas for $q_{\rm f} > 1$ (Right),
$\bar{N}_1$ is the only relevant saddle.

All these relevant saddles are considered to be of the Vilenkin type
since ${\rm Im} \bar{N}>0$, which corresponds to the wrong Wick rotation.
(See Section \ref{sec:geo} for the details.)
Thus
there is a possibility  of
the instability problem when one goes beyond
the mini-superspace model \cite{Feldbrugge:2017fcc,Feldbrugge:2017mbc}.



\subsection{Robin boundary condition}
\label{sec:RD-bc}

In order to
make the saddle points of the Hartle-Hawking type
relevant,
it was proposed \cite{DiTucci:2019dji,DiTucci:2019bui}
to impose the Robin boundary condition at $t=0$ as
\begin{alignat}{1}
  \label{Rbc}
  {3\pi^2\over N}\dot{q}(0)+\alpha+{q(0)\over\beta}=0 \ , 
\end{alignat}
where $\alpha$ and $\beta$ are complex parameters to be fixed later,
while keeping the Dirichlet boundary condition $q(1) =q_{\rm f}$ at $t=1$.
Note that the Robin boundary condition \eqref{Rbc}
interpolates the Dirichlet boundary condition
(with $q(0)=0$)
at $\beta=0$
and the Neumann boundary condition
(with $\dot{q}(0)=-\frac{N}{3\pi^2}\alpha$)
at $\beta=\infty$.

In fact, when
the Neumann boundary condition is used,
it is known \cite{Halliwell:1988ik}
that either Vilenkin saddles or Hartle-Hawking saddles
appear in the complex $N$ plane
depending on the sign of the initial Euclidean momentum
$- i \, \dot{q}(0) = i \frac{N}{3\pi^2} \alpha$. 
Furthermore, by choosing the value of the initial Euclidean momentum appropriately,
one can make the scale factor vanish at the initial time
for classical solutions.
However, quantum fluctuations of the scale factor diverge
due to
the uncertainty relation with fixed momentum.
This is the motivation for considering the Robin boundary
condition.\footnote{In order to make the Robin boundary condition
consistent with the variational principle,
one needs to add a boundary term \eqref{RBB} to the action by hand.
This is allowed from the viewpoint of defining Green's function,
which is taken throughout this paper.
Note, however, that this is not allowed from the viewpoint of
defining the wave function satisfying the Wheeler-DeWitt equation
based on the no-boundary proposal,
which requires the space-time manifold to have no boundary to the past.}

In order to obtain (\ref{Rbc}),
we add a boundary
term\footnote{This boundary term cannot be derived from a covariant term.
See Ref.~\cite{DiTucci:2019bui} for an alternative boundary condition, which
is compatible with a covariant boundary term.
See also Refs.~\cite{Ailiga:2023wzl,Ailiga:2024mmt} for applications
of the Robin boundary condition to the Gauss-Bonnet gravity.}
to the action \eqref{action-qn} given by
\begin{alignat}{1}
\tilde{S}_{\rm B}=
\alpha \, q(0)+{q(0)^2\over 2  \beta} \ .
\label{RBB}
\end{alignat}
Taking the variation of (\ref{RBB}) and combining it
with (\ref{delta-S-EH}),
we obtain
\begin{alignat}{1}
    - {3\pi^2\over N}
  \dot{q}(1)\delta q(1)
  +
  \left({3\pi^2\over N}\dot{q}(0)+\alpha+{q(0)\over\beta}\right)\delta q(0)
  & = 0 \ , 
\end{alignat}
which imposes
\eqref{Rbc}
on the classical solution since $q(0)$ is not fixed.

Similarly to the Dirichlet case,
we first solve
the classical equation of motion in \eqref{qEOM}
with the boundary conditions \eqref{Rbc} and $q(1)=q_{\rm f}$ as
\begin{alignat}{1}
  \bar{q}(t) &=
  N^2t^2 -  {\alpha\beta+q_{\rm f}-N^2\over3\pi^2\beta-N}Nt
  +{3\pi^2(q_{\rm f}-N^2)+N\alpha\over3\pi^2\beta-N}\beta \ ,
  \label{rsc}
\end{alignat}
and obtain the effective action
\begin{alignat}{1}
  S_{\rm eff}(N) &=
  \frac{\pi ^2 \left\{ N^4-6 N^2 (q_f-2)-3 q_f^2\right\}
    -\beta  \left\{ 6 \pi ^2 \alpha  \left(q_f-N^2\right)
    +12 \pi ^4 N \left(N^2-3 q_f+3\right)+\alpha ^2 N\right\} }
    {2 \left(N-3 \pi ^2 \beta \right) } \ .
    \label{action-eff-r}
\end{alignat}
It turns out that \eqref{action-eff-r} can be rewritten as
\begin{alignat}{1}
  S_{\rm eff}(N) &=
  \pi^2\left\{ {X^3\over2} - 3X(q_{\rm f}+A-2)
  - {3\over2X}(q_{\rm f}-A)^2\right\} + {\rm const.} \ ,
    \label{action-eff-r-2}
\end{alignat}
where we have defined
$X\equiv N-3\pi^2 \beta$ and $A\equiv 9 \pi^4 \beta^2- \alpha \beta$.
Since this has the same form as \eqref{eq:eff-S-DD}
up to the irrelevant constant term
with the correspondence $X \leftrightarrow N$ and $A \leftrightarrow q_{\rm i}$,
we can readily obtain four saddle points as
\begin{alignat}{1}
  \bar{N} &= c_1{\sqrt{q_{\rm f}-1}}
  + c_2{\sqrt{9\pi^4\beta^2 - \alpha\beta -1}} + 3\pi^2\beta \ ,
  \label{rnc}
\end{alignat}
where $c_1,\,c_2\in\{1,-1\}$.
Note that these saddle points reduce 
to those of the Dirichlet boundary conditions \eqref{DNcl}
with $q_{\rm i}=0$ for $\beta=0$.
Here we
define $\sqrt{9\pi^4\beta^2 - \alpha\beta -1}$
by analytic continuation with respect to $\beta$.
As in the Dirichlet case,
let us label them as $\bar{N}_1,\,\bar{N}_2,\,\bar{N}_3,\,\bar{N}_4$
corresponding to $(c_1,c_2) = (1,1)$, $(-1,1)$, $(-1,-1)$, $(1,-1)$,
respectively.

\begin{figure}[t]
   \centering
   \includegraphics[width=0.49\textwidth]{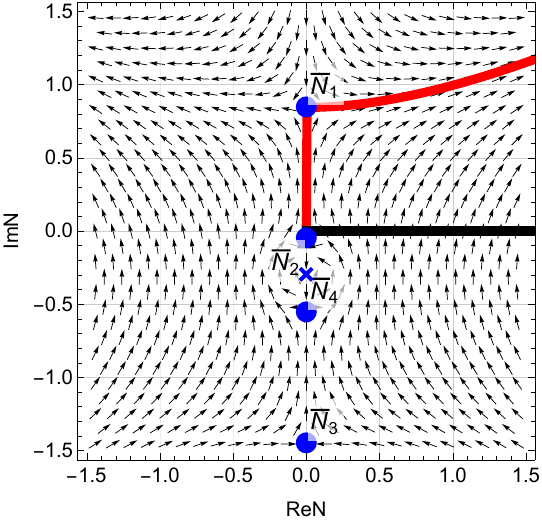}
   \includegraphics[width=0.49\textwidth]{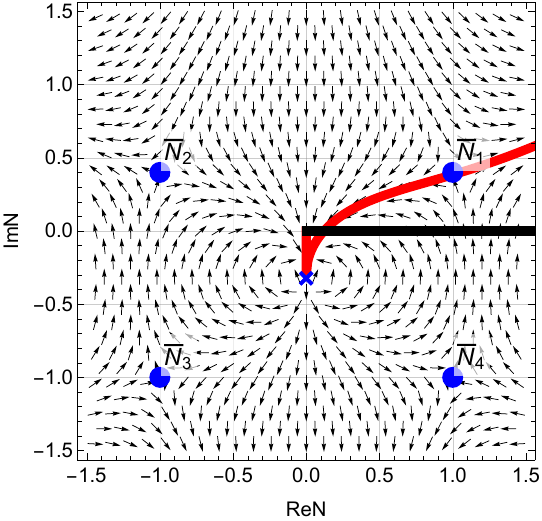}
   \caption{The flow diagram is shown similarly to 
Fig.~\ref{flowchart:D} in the case of
Robin boundary condition \eqref{Rbc} at $t=0$ with
$\alpha=-6\pi^2i$ and $\beta= -0.3 i\tilde{\beta}_{\rm c}$,
while imposing the Dirichlet boundary condition
$q_{\rm f}=0.8$ (Left) and $q_{\rm f}=10$ (Right),
respectively, at $t=1$.
%
}
   \label{flowchart:R}
\end{figure}

\begin{figure}[t]
   \centering
   \includegraphics[width=0.49\textwidth]{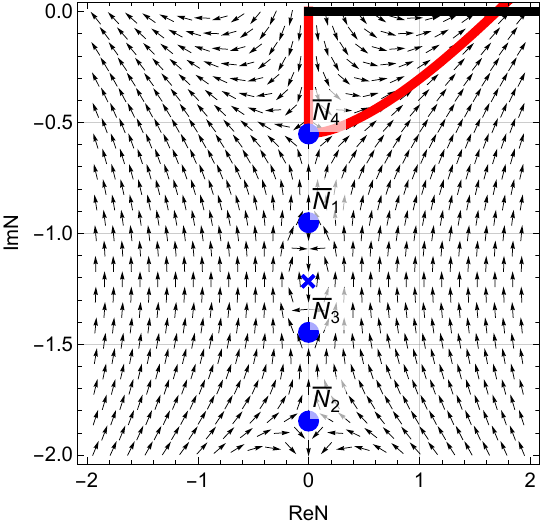}
   \includegraphics[width=0.49\textwidth]{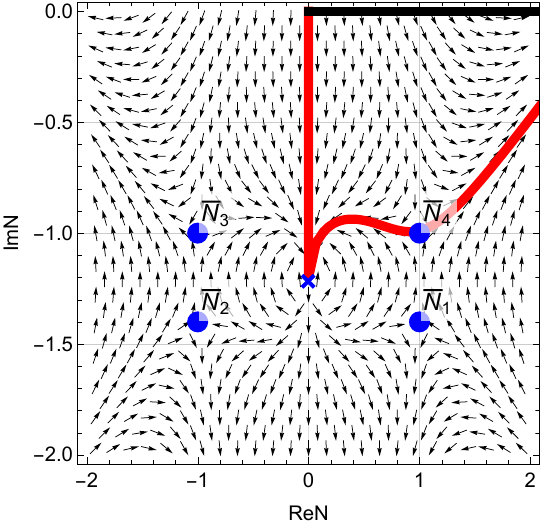}
   \caption{The flow diagram is shown similarly to 
Fig.~\ref{flowchart:D} in the case of
Robin boundary condition \eqref{Rbc} at $t=0$ with
$\alpha=-6\pi^2i$ and
$\beta=-1.2i\tilde{\beta}_{\rm c}$,
while imposing the Dirichlet boundary condition
$q_{\rm f}=0.8$ (Left) and $q_{\rm f}=10$ (Right),
respectively, at $t=1$.
}
   \label{flowchart:R2}
\end{figure}

In order to realize
the ``no boundary'' proposal,
we set $\alpha= - 6\pi^2i$ from now on for the reason which will be clear
shortly.\footnote{Similarly,
  if one chooses the opposite sign
  $\alpha= 6\pi^2i$, one finds that
the four saddle points become
$\bar{N}_{1,2}={i}\pm{\sqrt{q_{\rm f}-1}}$ and
$\bar{N}_{3,4}=-{i}\mp{\sqrt{q_{\rm f}-1}} +6\pi^2\beta$.
Thus the saddle points $\bar{N}_{1,2}$ are the same as in the Dirichlet case,
and in particular, they satisfy $\bar{N}^2 -2i \bar{N} - q_{\rm f}=0$.
Using this in \eqref{rsc} with $\alpha = 6 \pi^2 i$,
we obtain
$\bar{q}(t) = \bar{N}^2t^2 - 2i\bar{N}t$,
which is the same as \eqref{cl:Dqn} with $c_2=1$.
Thus, with this choice $\alpha = 6\pi^2 i$, we can make
$\bar{q}(0)=0$ for $\bar{N}_1$ and $\bar{N}_2$ for arbitrary $\beta$.
For the saddle points $\bar{N}_3$ and $\bar{N}_4$, one obtains
$\bar{q}(t) =
\bar{N}^2t^2 + 2(i + 6 \pi^2 \beta)\bar{N}t + 12\pi^2\beta(-i+3\pi^2\beta)$,
which do not satisfy $\bar{q}(0)=0$
except for $\beta=0$, $\frac{i}{3\pi^2}$.}
Plugging $\alpha=-6\pi^2i$ in (\ref{rnc}), we get
\begin{alignat}{1}
  \bar{N}&= c_1{\sqrt{q_{\rm f}-1}} + c_2 (i + 3 \pi^2 \beta) + 3 \pi ^2 \beta \ ,
\label{sol:robin}
\end{alignat}
which implies that
\begin{alignat}{1}
  \label{N12-beta}
\bar{N}_{1,2}&={i}\pm{\sqrt{q_{\rm f}-1}}+6\pi^2\beta \ ,\\
\bar{N}_{3,4}&=-{i}\mp{\sqrt{q_{\rm f}-1}} \ .
\label{N34-beta}
\end{alignat}
Thus the saddle points $\bar{N}_{3,4}$ are the same as in the Dirichlet case,
and in particular, they satisfy $\bar{N}^2 +2i \bar{N} - q_{\rm f}=0$.
Using this in \eqref{rsc} with $\alpha = - 6 \pi^2 i$,
we obtain 
\begin{alignat}{1}
  \bar{q}(t) &= \bar{N}^2t^2 + 2i\bar{N}t \ ,
  \label{cl:Rqn}
\end{alignat}
which is the same as \eqref{cl:Dqn} with $c_2=-1$.
Thus, with this choice $\alpha = -6\pi^2 i$, we can make
$\bar{q}(0)=0$ for $\bar{N}_{3,4}$ for arbitrary $\beta$ \cite{DiTucci:2019dji,DiTucci:2019bui}.
Similarly, for the saddle points $\bar{N}_{1,2}$,
we obtain 
\begin{alignat}{1}
  \bar{q}(t) &=
  \bar{N}^2t^2 - 2(i + 6 \pi^2 \beta)\bar{N}t + 12\pi^2\beta(i+3\pi^2\beta)
  \ ,
  \label{cl:Rqn-N12}
\end{alignat}
which do not satisfy $\bar{q}(0)=0$
except for $\beta=0$, $-\frac{i}{3\pi^2}$.

Below we focus on $\beta=-i\tilde{\beta}$
with $\tilde{\beta}\in \bbR_+$,
which shifts the saddle points
$\bar{N}_1$ and $\bar{N}_2$ downwards in the complex $N$ plane
as we increase $\tilde{\beta}$ while $\bar{N}_3$ and $\bar{N}_4$ remain the same,
as one can see from \eqref{N12-beta} and \eqref{N34-beta}.
Note, in particular, that $\bar{N}_1$ and $\bar{N}_2$ go below
$\bar{N}_4$ and $\bar{N}_3$, respectively, in the complex plane
at the critical value $\tilde{\beta}_{\rm c}\equiv {1\over3\pi^2}$.

Let us first discuss the case
$\tilde{\beta} < \tilde{\beta}_{\rm c}$.
In Fig.~\ref{flowchart:R}, we show the flow diagram
for the application of
the Picard-Lefschetz theory
to \eqref{action-eff-r}.
For both $q_{\rm f}=0.8$ (Left) and $q_{\rm f} =10$ (Right),
the relevant saddle point is $\bar{N}_1$,
which is of the Vilenkin type.\footnote{As we discuss in Section \ref{ap:arc},
  for $\tilde{\beta} \le \tilde{\beta}_{\rm c}/2$,
  there is a region $q_{\rm c}(\tilde{\beta}) \le q_{\rm f} \le 1$
  in which the relevant saddle points are $\bar{N}_{1}$ and $\bar{N}_{2}$.}


Next we consider 
the case $\tilde{\beta}>\tilde{\beta}_{\rm c}$.
The flow diagram is shown in Fig.~\ref{flowchart:R2}.
For both $q_{\rm f} <1$ (Left)
and
$q_{\rm f} >1$ (Right),
the relevant saddle point
is
$\bar{N}_{4}$,
which is of the Hartle-Hawking type,
and the corresponding scale factor satisfies
$\bar{q}(0)=0$ due to the chosen $\alpha = - 6 \pi^2 i$.

Thus the use of Robin boundary condition has enabled the realization
of the no-boundary proposal with the Hartle-Hawking saddle points
for the first time \cite{DiTucci:2019dji,DiTucci:2019bui}
within Lorentzian quantum gravity.
However, there is an issue
when one deforms the contour
since the end point $N=0$
of the integration region flows to another point, say $N=N^{(0)}$,
unlike the case of Dirichlet boundary condition,
where the end point $N=0$ coincides with the singularity
and hence does not flow. (See Fig.~\ref{flowchart:D}.)
This implies that the integration over the 
``arc'' between $N=0$ and $N=N^{(0)}$
has to be taken into account in the path integral
when one applies Cauchy's theorem.


\section{The arc problem for Robin boundary condition}
\label{ap:arc}

In this section, we discuss the issue concerning the arc contribution
that appears in the case of Robin boundary condition.
In particular, we
determine
the region of $q_{\rm f}$ in which
the arc contribution dominates over the contribution
from the thimble going through the relevant saddle point
in the $H^2\rightarrow0$ limit.

Let us note that the effective action \eqref{action-eff-r}
appears in the partition function as
$Z = \int dN e^{i S_{\rm eff}/H^2}$.
Since the relative weight $Z(N)\equiv |e^{i S_{\rm eff}(N) /H^2}|$ for each $N$
decreases along the flow from $N=0$,
the integral along the arc can be evaluated by the value at $N=0$ as
\begin{alignat}{1}
  Z(0)
=  \exp \left\{ -\frac{q_{\rm f}(q_{\rm f}-12\pi^2\tilde{\beta})}{2\tilde{\beta}H^2} +O(1)
  \right\}
       \label{action-eff-N-0}
\end{alignat}
in the $H^2\rightarrow0$ limit.
On the other hand, substituting $N=0$ in the classical solution \eqref{rsc},
one obtains $\bar{q}(t)=q_{\rm f}$, which
implies that there is no time-evolution.
Thus the dominance of the arc contribution
poses a serious problem in quantum cosmology.

\begin{figure}[t]
   \centering
   \includegraphics[width=0.8\textwidth]{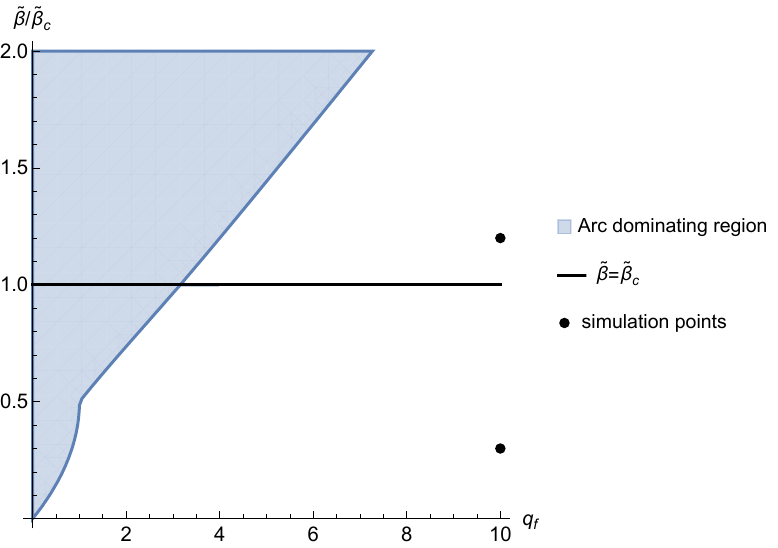}
   \caption{The shaded region represents
     the region of $q_{\rm f}$ and $\tilde{\beta}$
     in which the arc contribution that appears in the case of
     Robin boundary condition dominates over the contribution from
     the thimble going through 
     the relevant saddle point. The dots represent the parameter sets
     used for our simulations in Section \ref{sec:results}, which correspond
     to Fig.~\ref{flowchart:R} (Right) and Fig.~\ref{flowchart:R2} (Right),
     respectively.}
   \label{fig:arcregion}
\end{figure}

Let us first consider the case $\tilde{\beta}>\tilde{\beta}_{\rm c}$.
The relevant saddle point is
$\bar{N}_{4}$.
Plugging \eqref{N34-beta}
in the effective action \eqref{action-eff-r},
we obtain
\begin{equation}
  Z(\bar{N}_{4})
  =
  \left\{
  \begin{array}{ll}
    \exp \left\{ \frac{4\pi^2}{H^2} +O(1) \right\}  & \mbox{for $q_{\rm f}>1$} \ ,
       \label{action-eff-N3-N4}
    \\
    \exp \left\{ \frac{4 \pi ^2 }{H^2} \left(1-(1-q_{\rm f})^{3/2}\right) +O(1) \right\}
     & \mbox{for $q_{\rm f}<1$}    \ .
\end{array}
  \right.
\end{equation}
The condition for the dominance of the arc contribution is given by
$Z(0) > Z(\bar{N}_{4})$.
For $q_{\rm f}>1$, we obtain
\begin{alignat}{2}
  q_{\rm f}< q_{\rm c}(\tilde{\beta}) \equiv
  6\pi^2\tilde{\beta}+2\pi\sqrt{\tilde{\beta}(9\pi^2\tilde{\beta}-2)} \ .
\end{alignat}
Note that
$q_{\rm c}(\tilde{\beta}) > q_{\rm c}(\tilde{\beta}_{\rm c}) = 2 (1 + \frac{1}{\sqrt{3}})$
for $\tilde{\beta} >\tilde{\beta}_{\rm c}$.
For $q_{\rm f}<1$,
we obtain
\begin{alignat}{2}
  \tilde{\beta} > 
  \frac{q_{\rm f}^2}{4\pi^2(-2+2\sqrt{1-q_{\rm f}}+3q_{\rm f}-2\sqrt{q_{\rm f}-1}q_{\rm f})} \ .
\end{alignat}
Since the right-hand side, which we denote as $f(q_{\rm f})$,
decreases monotonically for $0<q_{\rm f}<1$, its maximum is given by
$\lim_{q_{\rm f}\rightarrow0}f(q_{\rm f})=\frac{1}{3\pi^2}=\tilde{\beta}_{\rm c}$.
Hence, this inequality is always satisfied.
Combining these two results,
we conclude that the contribution of the arc dominates
for $q_{\rm f} < q_{\rm c}(\tilde{\beta})$ in the $H^2\rightarrow0$ limit.


Similar analysis can be done for $\tilde{\beta}<\tilde{\beta}_{\rm c}$.
At the critical point $\tilde{\beta}=\tilde{\beta}_{\rm c}$,
the relevant saddle point for $q_{\rm f}>1$
switches from $\bar{N}_{4}$ to $\bar{N}_{1}$.
Therefore, the condition of our interest becomes $Z(0) > Z(\bar{N}_{1})$,
which gives
$q_{\rm f} < q_{\rm c}(\tilde{\beta}) \equiv 6 \pi ^2 \tilde{\beta}
+ 2 \pi \sqrt{ \tilde{\beta} (2 - 27 \pi^2 \tilde{\beta} + 108 \pi^4 \tilde{\beta}^2
  - 108 \pi ^6 \tilde{\beta}^3 )}$
for $\tilde{\beta}\ge  \tilde{\beta}_{\rm c}/2$.
At $\tilde{\beta}= \tilde{\beta}_{\rm c}/2$,
one finds that
the relevant saddle point $\bar{N}_{1}$ lies on the real axis
and the above expression gives $q_{\rm c}(\tilde{\beta}_{\rm c}/2)=1$.
Therefore, one has to consider the $q_{\rm f} < 1$ region for 
$\tilde{\beta}< \tilde{\beta}_{\rm c}/2$.
In this case, the arc contribution dominates when $\bar{N}_2$ lies
on the lower half-plane as one can see from Fig.~\ref{flowchart:R} (Left),
which gives
$q_{\rm f} < q_{\rm c}(\tilde{\beta}) \equiv 12 \pi^2 \tilde{\beta} (1 - 3 \pi ^2 \tilde{\beta})$.

Thus we obtain Fig.~\ref{fig:arcregion},
where we show the parameter region in which the arc problem occurs.
For instance, in
Fig.~\ref{flowchart:R} ($\tilde{\beta}=0.3\tilde{\beta}_{\rm c}$) and
Fig.~\ref{flowchart:R2} ($\tilde{\beta}=1.2\tilde{\beta}_{\rm c}$),
the arc problem occurs on the left ($q_{\rm f}=0.8$), but
not on the right ($q_{\rm f}=10$).

\section{Applying the GTM to quantum cosmology}
\label{sec:GTM-qc}

Since the integrand of the partition function \eqref{def-Z-mini-sup}
of quantum cosmology
is a pure phase factor,
it is difficult to apply
ordinary Monte Carlo methods
due to the sign problem.
In this section, we explain how to apply the GTM,
which is a promising method to overcome the sign problem.
See Appendix \ref{sec:GTM} for a brief review of the GTM.

\subsection{Discretized model for the Dirichlet boundary condition}
\label{sec:discrete-model-dirichlet}
Let us first
discuss how to discretize
the mini-superspace model in the case of
Dirichlet boundary conditions.
We start from the partition function \eqref{def-Z-mini-sup}
with the Dirichlet boundary conditions $q(0)=0$ and $q(1)=q_{\rm f}$.

We discretize the time coordinate $t$
with the step size $\epsilon$ satisfying $\ntime \epsilon = 1$
and define
$q_k= q(k \epsilon)$, where $k=0, \cdots , \ntime$,
and $p_k = p((k - \frac{1}{2})\epsilon)$, where $k=1, \cdots , \ntime$.
Then the partition function can be written as\footnote{The integration measure
for $q_k$ and $p_k$ is taken to be shift symmetric following Ref.~\cite{Halliwell:1988ik},
which is based on the standard canonical quantization procedure 
regarding $q_k$ and $p_k$ as the canonically conjugate variables.}
\begin{alignat}{2}
  Z &= \int dN\int\prod^{n-1}_{k=1} dq_k \int \prod^n_{k=1} dp_k\,e^{iS/H^2} \ ,
  \nonumber\\
  S &=
  \sum_{k=1}^{\ntime}\epsilon\left[
\frac{ \left(q_{k}-q_{k-1}\right)p_k}{\epsilon } +
\frac{N}{6 \pi ^2}
\left\{ p_k^2+36 \pi ^4 \left(1-
\frac{q_{k}+q_{k-1}}{2} \right)\right\} \right] \ ,
\end{alignat}
where we impose the Dirichlet boundary conditions
$q_0= 0$ and $q_{n}=q_{\rm f}$.
Integrating out $p_k$, we get
the effective theory for $q_k$ and $N$ as
\begin{alignat}{2}
Z &=  \int dN\, N^{-n/2}\int\prod^{n-1}_{k=1} dq_k \,e^{iS_{\rm eff}/H^2} \ ,
\label{discretized-Z-DD}
\\
S_{\rm eff} &=
6 \pi^2 \epsilon \sum_{k=1}^{\ntime}
\left\{ -\frac{1}{4N} \left({q_{k}-q_{k-1}\over\epsilon}\right)^2+ N \left(1- {q_k +q_{k-1}\over2} \right) \right \}  \ ,
\label{discretized-action-DD}
\end{alignat}
omitting an overall constant factor.

Taking the derivative of the effective
action $S_{\rm eff}$ with respect to $q_k$, we obtain
\begin{align}
  \frac{1}{\epsilon} \left(
  \frac{q_{k+1} - q_k }{\epsilon} - \frac{q_{k} - q_{k-1} }{\epsilon}
  \right) = 2 N^2  \quad \quad  \mbox{for $k=1, \cdots , n-1$} \ ,
  \label{qEOM-discrete}
\end{align}
which becomes the classical equation of motion \eqref{qEOM} for $q(t)$
in the $\epsilon \rightarrow 0$ limit correctly.



Note that $N=0$ is a singularity of the effective action \eqref{eq:eff-S-DD}
after integrating out $q(t)$,
which remains to be the case also for the discretized
model; see \eqref{action:eff-D-discrete}.
Therefore,
one cannot cross $N=0$ during the simulation\footnote{In the GTM, we
  deform the integration contour into the complex plane
  by solving the gradient flow \eqref{floweq}.
  When we say that $N$ cannot cross $N=0$,
  we actually mean the value of $N$ before the flow.}
based on the HMC algorithm, which is used in this work.
(See Appendix \ref{sec:GTM} for the details.)
Namely, if the initial value of $N$
is chosen to be positive,
it remains to be always positive.
Thus
the integration domain of $N$ is practically
restricted to
the $N > 0$ region.


\subsection{Discretized model for the Robin boundary condition}
\label{sec:discrete-model-robin}

As for the case of Robin boundary condition,
we derive the explicit form of 
the partition function
starting from
the path integral formalism in the phase space $q_k$ and $p_k$
generalizing the derivation given
in Ref.~\cite{Halliwell:1988ik}
for Dirichlet and Neumann boundary conditions.

Let us consider
the partition function \eqref{def-Z-mini-sup}
with the action \eqref{eq:S-q-p-N}
plus the boundary term \eqref{RBB}
as
\begin{alignat}{2}
Z &= \int dN\int {\cal D} q \int {\cal D} p\,e^{iS/H^2} \ ,
\\
S[p,q,N] &=
\int dt
\left[ \dot{q}p+{N\over 6\pi^2} \{ p^2+36\pi^4(1-q) \} \right]
+ \left( \alpha q(0)+{q(0)^2\over2\beta}\right)
\ ,
\end{alignat}
where we impose the Dirichlet boundary condition $q(1)=q_{\rm f}$ at the
final time.

As we did in the previous section, we 
discretize the time coordinate $t$ and obtain
\begin{alignat}{2}
  Z &=
  \int dN\int\prod^{n-1}_{k=0} dq_{k}
   \int \prod^n_{k=1} dp_k \, e^{iS/H^2} \ , \\
   S &=
     \sum_{k=1}^{\ntime}\epsilon
   \left[
\frac{ \left(q_{k}-q_{k-1}\right)p_k }{\epsilon }+
\frac{N}{6 \pi ^2}
\left\{ p_k^2+36 \pi ^4 \left(1-\frac{q_{k}+q_{k-1}}{2} \right)\right\}
\right]
+ \left( \alpha q_0+{q_0^2\over2\beta}\right)
\  ,
\end{alignat}
where we impose the Dirichlet boundary condition $q_{\ntime}=q_{\rm f}$
at the final time.
Integrating out $p_k$, we obtain
\begin{alignat}{2}
  Z &=  \int dN\, N^{-n/2}\int\prod^{n-1}_{k=0} dq_k \,e^{iS_{\rm eff}/H^2} \ ,
  \label{partition-fn-robin-q}
\\
S_{\rm eff} &=
  6 \pi^2 \epsilon
\sum_{k=1}^{\ntime}
\left\{ -\frac{1}{4N} \left({q_{k}-q_{k-1}\over\epsilon}\right)^2+ N \left(1- {q_k +q_{k-1}\over2} \right) \right \}
+ \left( \alpha q_0+{q_0^2\over2\beta}\right) 
  \ ,
\label{action-robin-q}
\end{alignat}
omitting an overall constant factor.

Taking the derivative of the effective action $S_{\rm eff}$
with respect to $q_k$ ($k=1, \cdots , n-1$),
we obtain the classical equation of motion \eqref{qEOM-discrete} as before.
Taking the derivative of
$S_{\rm eff}$
with respect to $q_0$, we obtain
\begin{align}
  \frac{3\pi^2}{N} \left(
  \frac{q_{1} - q_0 }{\epsilon} - N^2 \epsilon \right) + \alpha + \frac{q_0}{\beta}
  &= 0 \ .
  \label{Rbc-discrete}
\end{align}
Let us note here that
\begin{align}
  \frac{q_{1} - q_0 }{\epsilon} &\simeq
  q'\left(\frac{\epsilon}{2}\right) =   q'(0) + \frac{\epsilon}{2} \, q''(0)
    = q'(0) + N^2 \epsilon  \ .
\end{align}
Thus, in the $\epsilon \rightarrow 0$ limit,
\eqref{Rbc-discrete} reduces to
the Robin boundary condition \eqref{Rbc},
which is imposed on the classical solution.
As we explained in Section \ref{sec:RD-bc},
we set $\alpha = - 6 \pi^2 i$ and $\beta = - i \tilde{\beta}$
with $\tilde{\beta}\in \bbR_+$.


Let us note here that
the action \eqref{action-robin-q} is singular at $N=0$,
which causes a problem if the value of $N$ before the flow in the GTM
comes close to $N=0$.
In the case of Dirichlet boundary condition discussed in
Section \ref{sec:discrete-model-dirichlet},
the point $N=0$ is also
a singularity of the effective action \eqref{eq:eff-S-DD}
(See Fig.~\ref{flowchart:D}.),
and hence it is not approached during the simulation.
In the case of Robin boundary condition, however,
the singularity of the effective action \eqref{action-eff-r}
is shifted to $N_{\rm sing}=- 3 \pi^2 i \tilde{\beta}$
(See Figs.~\ref{flowchart:R} and \ref{flowchart:R2}.),
and hence $N=0$ may be approached.
In fact, this problem does not occur in practice
for $q_{\rm f}> q_{\rm c}(\tilde{\beta})$,
where $q_{\rm c}(\tilde{\beta})$ is obtained in Section \ref{ap:arc},
since in that region the contribution from the arc is suppressed compared with
the contribution from the thimble that goes through
the relevant saddle point.
Therefore the simulation is restricted practically
to the $N>0$ region before the flow,
and one cannot cross $N=0$ and go into the $N<0$ region.
The situation is similar to the case of Dirichlet boundary condition.
Hence the calculation corresponds to integrating over the thimble
associated with the relevant saddle with ${\rm Re}\bar{N}>0$
without including the arc contribution,
which is a good approximation
as far as $q_{\rm f} \gg q_{\rm c}(\tilde{\beta})$.
We restrict ourselves to this case in our simulation.

In Refs.~\cite{DiTucci:2019dji,DiTucci:2019bui},
it was proposed to extend the integration contour
to the whole real axis so that there is no contribution from the arc
in the whole parameter region.
If one wishes to
perform Monte Carlo simulation in this case,
one needs to
be able to go across the singular point $N=0$
in the action during the simulation.
This is possible
by shifting the integration contour of $N$
in the direction of the flow as
$N \mapsto N + i \delta$
with some small real constant $\delta$.
In addition to this issue, for $q_{\rm f}> q_{\rm c}(\tilde{\beta})$,
the probability of approaching the region ${\rm Re}N \sim 0$
is suppressed according to the discussion in Section \ref{ap:arc}.
Therefore, in order to be able to sample configurations
in the ${\rm Re}N>0$ and ${\rm Re}N<0$ regions
with equal weight,
one needs to use a more sophisticated version of the GTM
that makes it possible to sample configurations from
multiple thimbles \cite{Fukuma:2020fez,Fujisawa:2021hxh}.
We leave these cases for future investigations.

\section{Results of the simulations}
\label{sec:results}
In this section, we present the results of our simulation.
In particular, we compare the cases of 
the Dirichlet and Robin boundary conditions at the initial time,
and discuss how the relevant saddle point changes from
that of the Vilenkin type to that of the Hartle-Hawking type
in the case of Robin boundary condition as we change the parameter $\beta$.
For all the simulations in this work, the number of steps in time
is chosen to be $\ntime=10$. We have verified that the result does not change
significantly by increasing $\ntime$ to 20. 

\subsection{Dirichlet boundary condition}
\begin{figure}[t]
   \centering
   \includegraphics[width=0.49\textwidth]{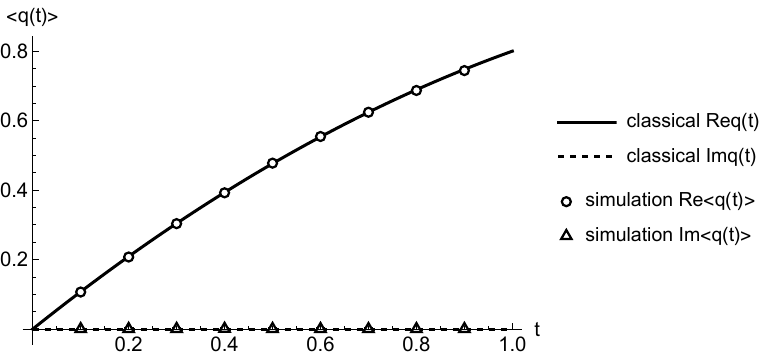}
   \includegraphics[width=0.49\textwidth]{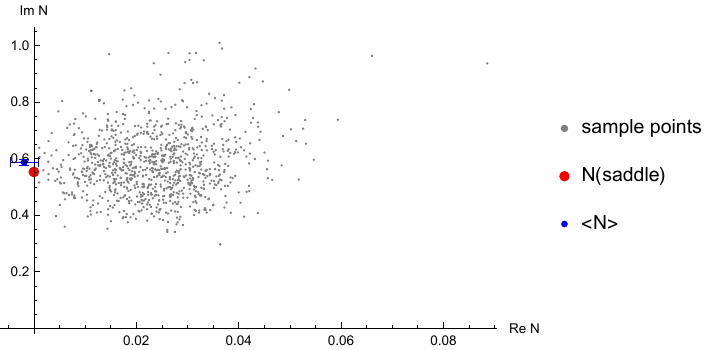}
   \includegraphics[width=0.49\textwidth]{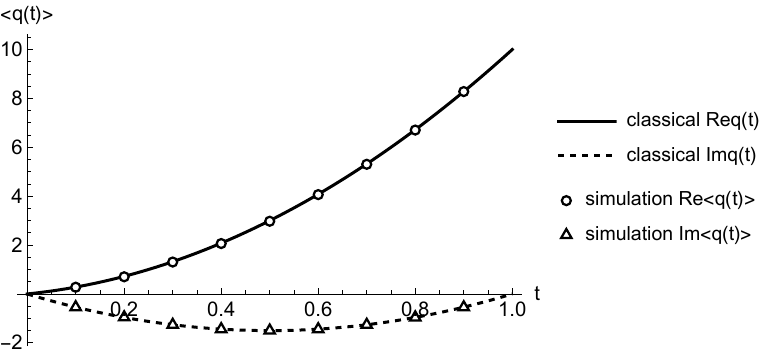}
   \includegraphics[width=0.49\textwidth]{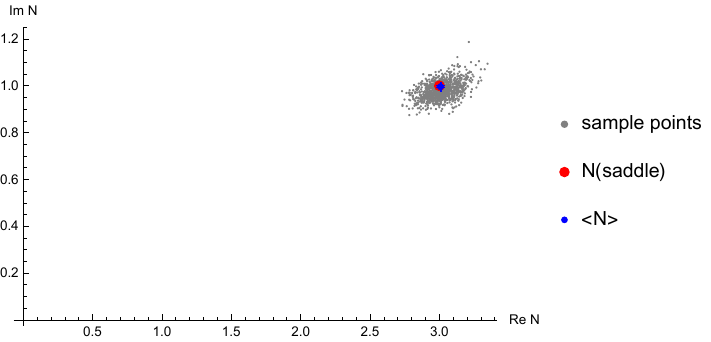}
   \caption{Simulation results are shown
     in the case of Dirichlet boundary condition for
$H^2=1$ with $q_{\rm f}=0.8$ (Top) and $q_{\rm f}=10$ (Bottom).
     (Left) The expectation values $\langle q(t)\rangle$ are plotted
     for the real (circles) and imaginary (triangles) parts separately.
     The classical solutions are shown for comparison
     by the solid and dashed lines for the real and imaginary parts, respectively.
     (Right) The gray dots represent the values of $N$ in the complex plane
     for each configuration obtained after the gradient flow.
     The blue circle represents
     $\langle N \rangle$,
     which is obtained by taking an average with a reweighting factor \eqref{rew},
     whereas the red circle represents the saddle point
     $\bar{N}_2$ in the Top-Right panel and
     $\bar{N}_1$ in the Bottom-Right panel,
     given by \eqref{DNcl-2} with $(c_1,c_2) = (-1,1)$ and $(1,1)$, respectively.
     The error bars for the expectation values are omitted
     when they are smaller than the data points.}
   \label{plot:D}
\end{figure}

First let us discuss the case of Dirichlet boundary condition.
In Fig.~\ref{plot:D} we plot the expectation values
$\langle q (t)\rangle$ for $t=k \epsilon$
($k=1 , \cdots , n-1$)
and $\langle N\rangle$ obtained
for $H^2=1$ with $q_{\rm f}=0.8$ (Top) and $q_{\rm f}=10$ (Bottom),
which correspond to the left and right panels of Fig.~\ref{flowchart:D}, respectively. 
We find good agreement
with the results\footnote{Let us recall that there are two relevant saddle
points in the $q_{\rm f}<1$ case represented by $(c_1,c_2) =(\pm 1,1)$
in \eqref{DNcl-2}.
Our results agree well with the results obtained for the dominant
saddle point $c_1=-1$ as expected.} of
the saddle-point analysis
\eqref{DNcl-2} and \eqref{cl:Dqn}
despite 
the fact that the chosen $H^2=1$ is not
small enough to suppress
quantum corrections as we discuss in Section \ref{sec:quantum-corrections}.

In particular, when $q_{\rm f}<1$ (the upper panels),
$\langle N \rangle$ is purely imaginary and 
$\langle q(t) \rangle$ is purely real, 
indicating that the
geometry dominating the path integral in this case is
Euclidean,
and it is actually a part of a hemisphere as we discuss in Section \ref{sec:geo}.
The fact ${\rm Im} \langle N \rangle > 0$ implies that
the relevant saddle is that of the Vilenkin type as expected.
On the other hand,
when $q_{\rm f}>1$ (the lower panels), 
the real part of $\langle N \rangle$ 
becomes positive and $\langle q(t) \rangle$ acquires an imaginary part.
Thus the dominant geometry becomes complex in this case.
However, this is simply due to the chosen gauge
$dN(t)/dt=0$
for the time-reparametrization invariance,
and it actually represents a part of de Sitter space capped off by
a Euclidean hemisphere as we discuss in Section \ref{sec:geo}.
Thus the origin of real time can be identified
with the Stokes phenomenon at the critical value $q_{\rm f}=1$
as emphasized in Ref.~\cite{Lehners:2021jmv};
the relevant saddle points
$\bar{N}_1$ and $\bar{N}_2$ on the imaginary axis for $q_{\rm f}<1$
merge into one at $\bar{N}_1 = \bar{N}_2 = i$ for $q_{\rm f}=1$
and split into two for $q_{\rm f}>1$ acquiring the real part
as we have seen in Fig.~\ref{flowchart:D}.


In the right panels, the gray dots 
represent the values of $N$ calculated for
sampled configurations
after the gradient flow in the GTM.
Note that they are not necessarily distributed around the expectation
value $\langle N\rangle$,
which is obtained by taking an average with a reweighting
factor \eqref{exp-avg}.




\subsection{Robin boundary condition}

\begin{figure}[t]
   \centering
   \includegraphics[width=0.49\textwidth]{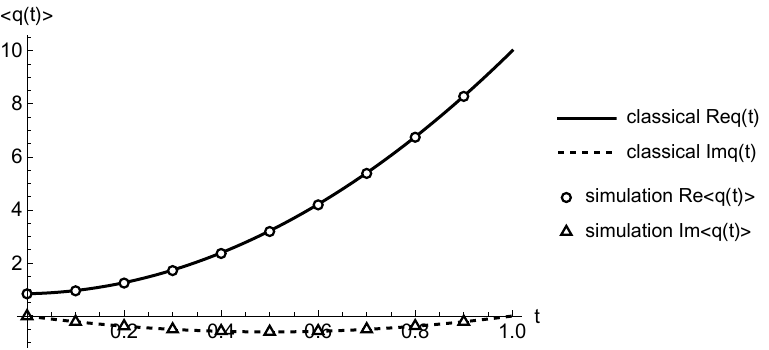}
   \includegraphics[width=0.49\textwidth]{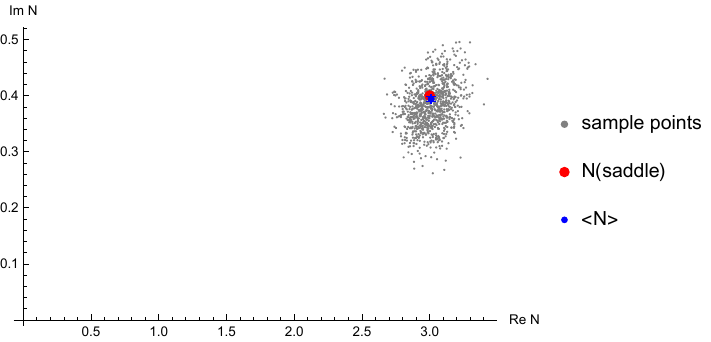}
   \includegraphics[width=0.49\textwidth]{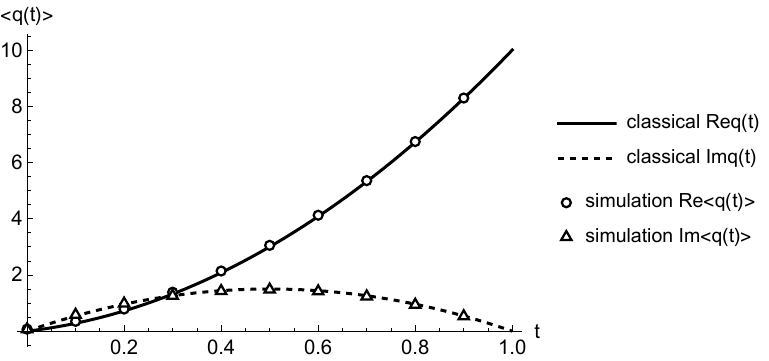}
   \includegraphics[width=0.49\textwidth]{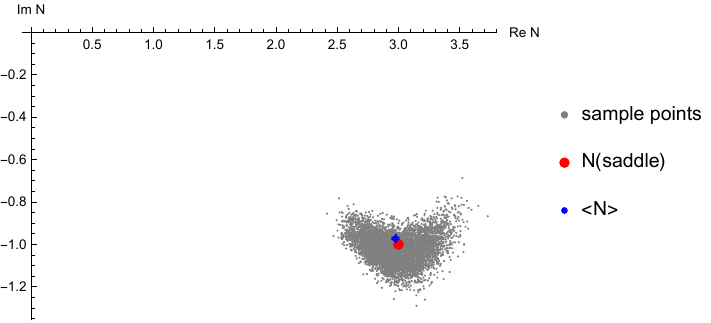}
   \caption{Similar plots to Fig.~\ref{plot:D}
in the case of Robin boundary condition for 
$H^2=1$ and $q_{\rm f}=10$
with $\tilde{\beta}=0.3\tilde{\beta}_{\rm c}$ (Top)
and $\tilde{\beta}=1.2\tilde{\beta}_{\rm c}$ (Bottom).
The red point represents
$\bar{N}_1$ in the Top-Right panel and
$\bar{N}_4$ in the Bottom-Right panel,
given by \eqref{N12-beta} and \eqref{N34-beta}, respectively.
}
   \label{plot:Rsb}
\end{figure}

Next we show our results in the case of Robin boundary condition.
In Fig.~\ref{plot:Rsb} (Top),
we plot our results for $H^2=1$
with $\tilde{\beta}=0.3\tilde{\beta}_{\rm c}$ and $q_{\rm f}=10$,
which corresponds to
Fig.~\ref{flowchart:R} (Right).
We find good agreement with the results obtained by the
saddle-point analysis \eqref{N12-beta}
and \eqref{cl:Rqn-N12}.
The results are similar to the corresponding results for the
Dirichlet boundary condition shown in the
lower panel of Fig.~\ref{plot:D} as expected.
In particular, 
${\rm Im} \langle N \rangle > 0$ implies that the relevant saddle is 
that of 
the Vilenkin type
in this case.
Note also that $\langle q(0) \rangle \neq 0$
due to our choice $\alpha = - 6 \pi^2 i$.
(See Eq.~\eqref{cl:Rqn-N12}.)



\begin{figure}[t]
   \centering
   \includegraphics[width=0.49\textwidth]{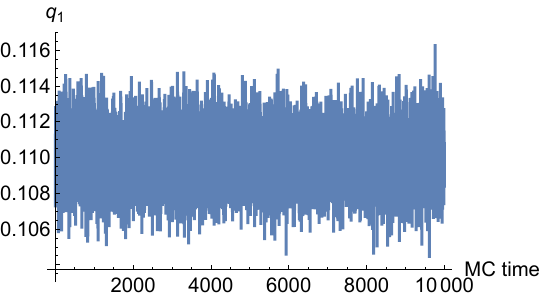}
   \includegraphics[width=0.49\textwidth]{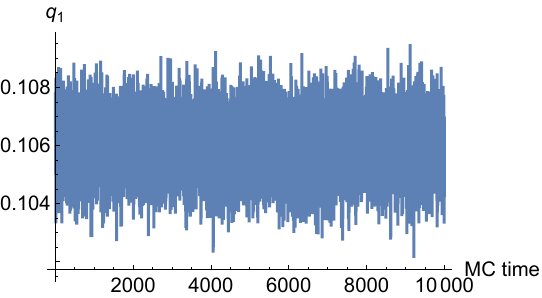}
   \includegraphics[width=0.49\textwidth]{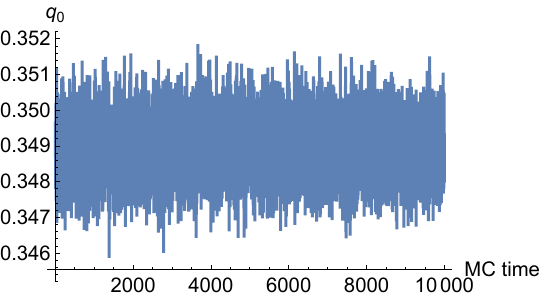}
   \includegraphics[width=0.49\textwidth]{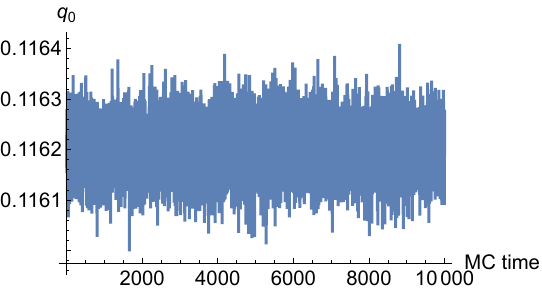}
   \caption{The Monte Carlo history of the smallest $q_k=q(k \epsilon)$
     before the gradient flow in the GTM is shown.
     The horizontal axis refers to the number of
     sampled configurations.
     In the upper panels, we show the history of $q_1$ in the Dirichlet case
     for $H^2=1$ with $q_{\rm f}=0.8$ (Left) and $q_{\rm f}=10$ (Right).
     In the lower panels, we show the history of $q_0$ in the Robin case
     for $H^2=1$ and $q_{\rm f}=10$
     with $\tilde{\beta}=0.3\tilde{\beta}_{\rm c}$ (Left)
     and $\tilde{\beta}=1.2\tilde{\beta}_{\rm c}$ (Right).
}
   \label{plot:qh}
\end{figure}

In Fig.~\ref{plot:Rsb} (Bottom), we show our results for $H^2=1$
with $\tilde{\beta}=1.2\tilde{\beta}_{\rm c}$ and $q_{\rm f}=10$,
which corresponds to
Fig.~\ref{flowchart:R2} (Right).
We find good agreement with the results obtained by the
saddle-point analysis \eqref{N34-beta} and \eqref{cl:Rqn}.
In particular, ${\rm Im} \langle N \rangle < 0$ implies that
the relevant saddle is that of the Hartle-Hawking type.
Note also that $\langle q(0) \rangle \approx 0$, which
confirms that the ``no boundary'' proposal is realized 
in the sense of expectation values due to the choice $\alpha = - 6 \pi^2 i$.
On the other hand,
the real part of $\langle N \rangle$ 
becomes positive and $\langle q(t) \rangle$ acquires an imaginary part.
We will see in Section \ref{sec:geo} that
the real geometry one can read off in this case is
a part of de Sitter space capped off by a Euclidean hemisphere 
as in the Dirichlet case although one has to make a Wick rotation in the
opposite orientation.



In passing,
let us also comment on the issue raised in Ref.~\cite{Jia:2022nda}
concerning the change of variables \eqref{q-a-relation}.
In order for the variable $q(t)$ to correspond to the scale factor squared,
it is necessary that the path integral over $q(t)$ is restricted
to the $q(t)\ge 0$ region.
In Fig.~\ref{plot:qh}, we present the history of the smallest $q_k=q(k\epsilon)$
[$k=1$ for the Dirichlet case and $k=0$ for the Robin case]
before the gradient flow in the GTM.
We find that only positive $q_k$ appears during the simulation
despite the fact that
$\langle q_k \rangle$ is close to zero when the no-boundary proposal is realized.
Thus the results obtained by our simulation correspond practically to
restricting the path integral to the $q(t)\ge 0$ region.\footnote{Even if
this turns out not to be the case, for instance in the case of large $H^2$,
we can make a change of variables $q(t)=e^{\sigma(t)}$ with $\sigma(t)\in \bbR$
and adopt a shift-invariant measure for $\sigma(t)$, which
would exclude the \(q(t)<0\) region completely.
Exploring such a model, which is inequivalent to the original one,
is left for future investigations.}


\subsection{Comparison with the perturbative expansion in $H^2$}
\label{sec:quantum-corrections}

In all the simulation results obtained above with $H^2=1$,
the expectation values turned out to be
close to the results of the saddle-point analysis.
In this section, we calculate the quantum corrections
to the saddle-point results
by the perturbative expansion with respect to $H^2$
and demonstrate that the perturbation theory
breaks down at $H^2 \sim 0.1$.
This implies that the simulation point $H^2=1$ actually lies in the
nonperturbative regime, where perturbation theory is not applicable.
We also perform additional simulations with $H^2=0.3$, $0.1$, $0.03$
in order to see the consistency with the perturbation theory.
Here we focus on the case of Robin boundary condition
with $\tilde{\beta}=1.2\tilde{\beta}_{\rm c}$ and $q_{\rm f}=10$.

Since our simulation uses $\epsilon=0.1$ for discretizing the time $t$,
we have to consider the perturbative expansion in $H^2$
for finite $\epsilon$ in order to make a precise comparison.
For that we
start from the discretized partition function \eqref{partition-fn-robin-q}.
Integrating out $q_k$, we arrive at (See Appendix \ref{sec:eff-action-epsilon}
for derivation.)
\begin{alignat}{2}
Z &=\int dN\,e^{i S_{\rm eff}[N]/H^2} \ ,
\\
S_{\rm eff}[N]&= -\frac{\pi ^2 }{2}\bigg(
 \epsilon^2 N^3
+\frac{ \beta  \left(6 \pi ^2 \alpha  \left(q_f-N^2\right)+12 \pi ^4 N \left(N^2-3 q_f+3\right)+\alpha ^2 N\right)}{\pi ^2 \left(N-3 \pi ^2 \beta \right)}\nonumber\\
&-\frac{ N^4-6 N^2 (q_f-2)-3 q_f^2}{N-3 \pi ^2 \beta }\bigg)
+ {i \over2} H^2 \log \left(N-3\pi^2\beta\right) \ ,
\label{action:eff-r}
\end{alignat}
up to an irrelevant constant.
Thus the finite $\epsilon$ effect appears only in the first term.


\begin{figure}[t]
   \centering
   \includegraphics[width=0.7\textwidth]{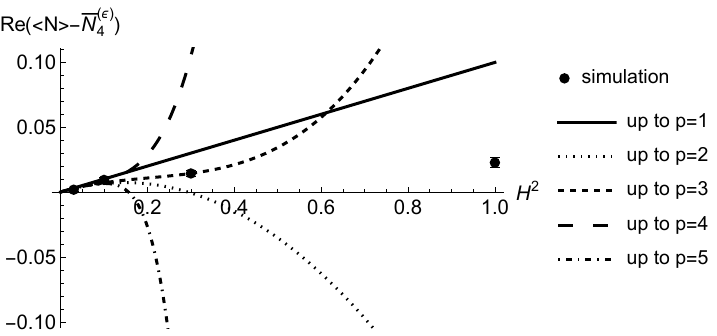}
   \includegraphics[width=0.7\textwidth]{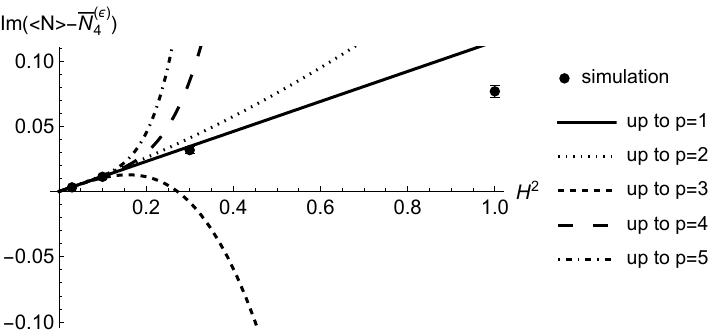}
   \caption{The expectation value $\langle N \rangle$
          is plotted against $H^2$
     for the real part (Left) and the imaginary part (Right), respectively,
     in the case of Robin boundary condition for
     various $H^2=0.03,\,0.1,\,0.3,\,1$
     with $\tilde{\beta}=1.2\tilde{\beta}_{\rm c}$ and $q_{\rm f}=10$.
     The lines represent the results \eqref{Nav-sim}
     obtained by perturbation theory
     up to the $p\,$-th order ($p=1 , \cdots , 5$) in $H^2$ with $\epsilon=0.1$.
}
   \label{plot:h}
\end{figure}

We obtain the solution to the saddle-point equation derived from
\eqref{action:eff-r} in the form of
the $H^2$ expansion as 
\begin{alignat}{2}
  N^{(0)} = \Bar{N}_4^{(\epsilon)} + \sum_{p=1}^{\infty}
  a_{p} \,  H^{2p} \ ,
\end{alignat}
around the relevant saddle point $\Bar{N}_4^{(\epsilon)}$
corresponding to
$\Bar{N}_4=-i+\sqrt{q_{\rm f}-1}$ in \eqref{N34-beta}
in the $\epsilon\rightarrow 0$ limit.
Then we substitute $N = N^{(0)} + \varphi$ in
the action and calculate
\begin{alignat}{2}
  \langle \varphi \rangle&={1\over Z}\int^\infty_{-\infty}
  d\varphi \,e^{i S_{\rm eff}[N^{(0)}+ \varphi]/H^2}
\end{alignat}
perturbatively, which yields
\begin{alignat}{2}
  \langle N\rangle = N^{(0)} + \langle \varphi \rangle = 
  \Bar{N}_4^{(\epsilon)}
  +   \sum_{p=1}^{\infty} \tilde{a}_{p} \, H^{2p} \ .
  \label{Nav-expand}
  %
\end{alignat}
For
$\epsilon = 0.1$ with $\tilde{\beta}=1.2\tilde{\beta}_{\rm c}$ and $q_{\rm f}=10$,
we obtain
\begin{align}
  {\rm Re} \langle N\rangle & =
  2.95266 + 0.099668 H^2 - 0.340375 H^4 + 0.555885 H^6 + 11.1378 H^8-172.735 H^{10}\ , \nn \\
  {\rm Im} \langle N\rangle & =
  -1.04925 + 0.115002 H^2 + 0.0718019 H^4 - 1.82104 H^6 + 11.4782 H^8 +47.0957 H^{10}\ ,
\label{Nav-sim}
\end{align}
up to the fifth order in $H^2$.

In Fig.~\ref{plot:h},
we plot
the expectation value $\langle N \rangle$
for $H^2=0.03,\,0.1,\,0.3,\,1$.
Our results
are in good agreement with the perturbative result
\eqref{Nav-sim}
for $H^2 \lesssim 0.1$, which confirms the validity of our simulations.
On the other hand, for $H^2 \gtrsim 0.1$,
the perturbative results show diverging behaviors as the order $p$
is increased,
which
implies
that the perturbation theory breaks down
around $H^2 \sim 0.1$.
Thus
$H^2=1$
used in our simulation
for Figs.~\ref{plot:D} and \ref{plot:Rsb}
lies in a regime in which perturbation theory is no more applicable.\footnote{The analysis
in this section corresponds to
our Monte Carlo results in the Bottom-Right panel of Fig.~\ref{plot:Rsb},
which are obtained for the Robin boundary condition
at the initial time
with $H^2=1$, $q_{\rm f}=10$ and $\tilde{\beta}=1.2\tilde{\beta}_{\rm c}$.
Note, however, that the red circle in Fig.~\ref{plot:Rsb} represents the saddle point
for $\epsilon = 0$ unlike the saddle point $\Bar{N}_4^{(\epsilon)}$
used in Fig.~\ref{plot:h}.
The smallness of quantum corrections even in the nonperturbative regime
may be attributed to the simplicity of this model, where quantum corrections
are caused only by fluctuations of the lapse $N$.
Note also that the radius of convergence
for the expansion \eqref{Nav-sim} with respect to $1/H^2$ is likely to be zero
based on the ratio test, which is not surprising for perturbative expansions.}
This demonstrates the capability of our method to
perform full quantum
calculations in quantum cosmology.

\section{Extracting real geometry from complex geometry}
\label{sec:geo}


In this section, we discuss
the metric of the space-time \eqref{metric-q-t}
for the relevant saddle points.
As we have seen in Sections \ref{sec:DD-bc} and \ref{sec:RD-bc},
for $q_{\rm f}<1$, $\bar{N}$ becomes purely imaginary
and the metric becomes real with the Euclidean signature.
On the other hand,
for $q_{\rm f}>1$, $\bar{N}$ and hence the metric becomes complex.
This has something to do with
the way we fix the ``gauge'' for
the time-reparametrization invariance of
the action \eqref{action-EH}.
Here we discuss
how to
read off the real geometry from the complex geometry at the saddle point,
refining the previous discussions on this issue.
(See Section III of Ref.~\cite{DiTucci:2019bui}, for instance.)

Note first that
the action \eqref{action-EH}
is invariant under the time-reparametrization given by
\begin{align}
  q(t) \mapsto \tilde{q}(t) \ , \quad 
  N(t) \mapsto \tilde{N}(t) \ ,
  \label{eq:time-rep-transf}
\end{align}  
where $\tilde{q}(t)$ and $\tilde{N}(t)$
are defined by
\begin{align}
\tilde{q}(f(t)) = q(t) \ , \quad
\tilde{N}(f(t)) \frac{df(t)}{dt} = N(t) \ ,
\label{eq:time-rep-def}
\end{align}  
for an arbitrary real function $f(t)$
with $df(t)/dt > 0$
so that $\tilde{N}(t)\ge 0$.
We also require that $f(0)=0$ and $f(1)=1$
so that the end points remain the same.
We have gauge-fixed this symmetry with the condition $dN(t)/dt=0$,
which resulted in 
the gauge-fixed action \eqref{eq:S-q-p-N-0}.

When we search for saddle points of the action \eqref{action-EH},
we have to
complexify $q(t)$ and $N(t)$.
Correspondingly, let us complexify the time-coordinate
$t \mapsto \tau(t) \in \bbC$,
where
$\tau(0)=\tau_{\rm i}$ and $\tau(1)=\tau_{\rm f}$.
This defines a contour in the complex-$\tau$ plane with the
end points $\tau_{\rm i}$ and $\tau_{\rm f}$,
which may be regarded as a generalization of the Wick rotation.
In fact, the \emph{value} of the action is invariant
under the replacement
\begin{align}
  t  \mapsto \tau \ , \quad
  q(t) \mapsto \tilde{q}(\tau) \ , \quad 
  N(t) \mapsto \tilde{N}(\tau) \ ,
  \label{eq:time-rep-transf-complex}
\end{align}  
where $\tilde{q}(\tau)$ and $\tilde{N}(\tau)$
are defined by
\begin{align}
\tilde{q}(\tau(t)) = q(t) \ , \quad
\tilde{N}(\tau(t)) \frac{d\tau}{dt} = N(t) \ .
\label{eq:time-rep-def-complex}
\end{align}  
Note that this is \emph{not} a symmetry
unlike the time-reparametrization\footnote{The deformed theory still has
  a symmetry under the time-reparametrization, which amounts to
  the replacement
  $\tau(t) \mapsto \tau(f(t))$ in \eqref{eq:time-rep-def-complex}.}
since the \emph{form} of the action is changed
by deforming the integration contour $t \mapsto \tau$.
In particular,
one can make the lapse function $\tilde{N}(\tau)=1$ by
choosing $\tau(t)$ to be the solution to
$d\tau / dt = N(t)$.

Next we point out that
the value of the action
in the
deformed
theory is invariant under
the change
of the $\tau$-integration contour
with the end points $\tau_{\rm i}$ and $\tau_{\rm f}$ fixed
due to Cauchy's theorem as far as
$q(\tau)$ and $N(\tau)$ are defined by analytic continuation.
Note that this invariance is different
from the invariance under \eqref{eq:time-rep-transf-complex}
for a different choice of $\tau(t)$ with the same end points.

Finally we use the deformation \eqref{eq:time-rep-transf-complex}
backwards to get back
to the original real-time theory.
Using this chain of transformations, one can
transform a saddle-point configuration into an equivalent configuration,
which cannot be obtained in the $dN(t)/dt=0$ gauge.
In particular, this enables us to read off
the real geometry
from the complex geometry at the saddle point as we see below.


Let us first discuss the case of Dirichlet boundary condition.
The relevant saddle points $\bar{N}$ are given by \eqref{DNcl-2} with $c_2=1$,
which is of the Vilenkin type; namely
${\rm Im}\bar{N}>0$,
with
$c_1=\pm 1$ for $q_{\rm f}<1$ and $c_1=1$ for $q_{\rm f}>1$.
The classical solution $\bar{q}(t)$
for the scale factor is given by \eqref{cl:Dqn}.
Below we restrict ourselves to the dominant saddle point ($c_1= 1$)
with ${\rm Im}\bar{N} < 1$ for $q_{\rm f}<1$.

Let us first
make a generalized Wick rotation
\begin{align}
  \tau = \bar{N} t   \ ,
\label{eq:gen-Wick-rot}
\end{align}
which transforms the saddle-point configuration into
\begin{align}
  \bar{\tilde{q}}(\tau) &= \bar{q}\left(\frac{\tau}{\bar{N}}\right) =
    \tau^2 - 2 i \tau
  \ , \\
  \bar{\tilde{N}}(\tau) &= \frac{1}{\bar{N}} \, 
  \bar{N}\! \left(\frac{\tau}{\bar{N}}\right) = 1 \ ,
\label{eq:time-rep-def-complex-D}
\end{align}  
where we have used \eqref{eq:time-rep-def-complex}.
Thus the lapse function becomes unity.

Next we change the $\tau$-integration contour as 
\begin{align}
  \tau(t) =
  \left\{
\begin{array}{ll}
i t & \mbox{~for~}0 \le t \le |{\rm Im} \bar{N}| \ , \\
i  + (t-1)  &
\mbox{~for~}1 < t \le  1+ {\rm Re} \bar{N} \ ,
\end{array}\right.
\label{t-tau-I}
\end{align}
which makes the classical solution $\bar{\tilde{q}}(\tau)$ real
\begin{align}
  \bar{\tilde{q}}(\tau(t)) & =
  \left\{
\begin{array}{ll}
- t^2+2t  & \mbox{~for~} 0\le t \le |{\rm Im} \bar{N}| \ , \\
t^2-2t +2  & \mbox{~for~} 1 < t \le 1+ {\rm Re} \bar{N} \ ,
\end{array}\right.
\label{q-tau-I-II}
\end{align}
along the new contour.
For $q_{\rm f}<1$, we have $|{\rm Im} \bar{N}|<1$,
${\rm Re} \bar{N}=0$ and the second lines
in \eqref{t-tau-I} and \eqref{q-tau-I-II}
do not exist,
whereas
for $q_{\rm f}>1$,
we have $|{\rm Im} \bar{N}|= 1$, ${\rm Re} \bar{N} > 0$ and 
the second lines exist.

Finally we
use the deformation \eqref{eq:time-rep-transf-complex}
backwards to get back
to the original real-time theory.
The saddle-point configuration one obtains in this
way\footnote{The range of time $t$ is $t =\in [0, T]$,
where $T={\rm Re} \bar{N} + {\rm Im} \bar{N}$.
In order to adjust the range to the original theory,
one can rescale the time as $t \mapsto \tilde{t}=t/T$.}
is $\bar{q}(t) = \bar{\tilde{q}}(\tau(t))$, which is given by \eqref{q-tau-I-II},
with the lapse function given by
\begin{align}
  \bar{N}(t) & =
  \left\{
\begin{array}{ll}
 i  & \mbox{~for~} 0\le t \le |{\rm Im} \bar{N}| \ , \\
 1  & \mbox{~for~} 1 < t \le 1+ {\rm Re} \bar{N} \ ,
\end{array}\right.
\label{N-I-II-t}
\end{align}
where we have used \eqref{eq:time-rep-def-complex} again.
Note that \eqref{N-I-II-t}
cannot be obtained in the $dN(t)/dt=0$ gauge.
The first line in \eqref{N-I-II-t} corresponds
the Euclidean regime, where the lapse function becomes purely imaginary,
while the second line in \eqref{N-I-II-t}, which exists for $q_{\rm f}>1$,
corresponds to the Lorentzian regime, where the lapse function becomes real.
Thus we can read off the real geometry at the saddle point,
which consists of the Euclidean regime at early times and
the Lorentzian regime at later times.


Let us define the proper time $\tau_{\rm p}$ by
making a time-reparametrization as
\begin{alignat}{1}
  {dt \over d \tau_{\rm p}} = \sqrt{\bar{q}(t)}
  \label{cov:Dsq}
\end{alignat}
so that the metric of the space-time \eqref{metric-q-t} becomes
\begin{eqnarray}
  ds^2 =  \pm  d\tau_{\rm p}^2 + \bar{q}(t(\tau_{\rm p})) \, d \Omega_3{}^2 \ ,
\label{metric-q-t-proper}
\end{eqnarray}
where the symbol $\pm$ should be $(+)$ for the Euclidean regime
and $(-)$ for the Lorentzian regime.
By solving the differential equation \eqref{cov:Dsq},
we obtain
\begin{align}
  t(\tau_{\rm p}) =
  \left\{
\begin{array}{ll}
  1 - \cos \tau_{\rm p}  &  \mbox{~for~}0\le \tau_{\rm p} \le \cos^{-1} (1-|{\rm Im} \bar{N}|)  \ , \\
  1 + \sinh (\tau_{\rm p} - \frac{\pi}{2} ) &
  \mbox{~for~} \frac{\pi}{2} < \tau_{\rm p}
  \le \frac{\pi}{2}+ \sinh^{-1} ({\rm Re} \bar{N})  \ ,
\end{array}\right.
\label{tau-t-proper}
\end{align}
which makes the classical solution $\bar{q}(t(\tau_{\rm p}))$ as
\begin{align}
  \bar{q}(t(\tau_{\rm p})) & =
  \left\{
\begin{array}{ll}
\sin ^2  \tau_{\rm p}  & \mbox{~for~}0\le \tau_{\rm p} \le \cos^{-1} (1-|{\rm Im} \bar{N}|)  \ , \\
\cosh ^2 (\tau_{\rm p} - \frac{\pi}{2})
& \mbox{~for~} \frac{\pi}{2} <
\tau_{\rm p}  \le \frac{\pi}{2} + \sinh^{-1} ({\rm Re} \bar{N})  \ .
\end{array}\right.
\label{q-tau-proper}
\end{align}
Thus for $q_{\rm f} < 1$,
the geometry of the saddle-point configuration is
a part of a hemisphere,
whereas for $q_{\rm f} > 1$,
it becomes a part of the de Sitter space
capped off by
a hemisphere.

In the case of
Robin boundary condition
for $\tilde{\beta} >\tilde{\beta}_{\rm c}$,
the relevant saddle point
is given by $\bar{N}_4$ in \eqref{N34-beta},
which is of the Hartle-Hawking type (${\rm Im}\bar{N}<0$).
The classical solution $\bar{q}(t)$
for the scale factor is given by \eqref{cl:Rqn}.
We make a generalized Wick rotation \eqref{eq:gen-Wick-rot},
and change the $\tau$-integration contour as 
\begin{align}
  \tau=
  \left\{
\begin{array}{ll}
- i t & \mbox{~for~} 0\le t \le |{\rm Im} \bar{N}| \ , \\
- i + (t-1)  &
\mbox{~for~} 1 < t \le  1+ {\rm Re} \bar{N}  \ ,
\end{array}\right.
\label{t-tau-I-Robin}
\end{align}
which makes the classical solution $\bar{\tilde{q}}(\tau)$
in the same form \eqref{q-tau-I-II} as in the Dirichlet case.
Thus the only difference from the case of Dirichlet boundary condition
is the orientation of the Wick rotation as one can see from
\eqref{t-tau-I} and \eqref{t-tau-I-Robin}.

While we have shown clearly how to extract real geometries from the complex saddle points
in the case at hand,
the saddle point geometries in more general setups
are inherently complex and cannot be always rendered real as described here.
The complex nature of such saddle points
has been explored in the recent literature,
notably in Ref.~\cite{Witten:2021nzp}
and more specifically in Refs.~\cite{Jonas:2022uqb,Hertog:2023vot}
in the context of the no-boundary proposal.

\section{Summary and discussions}
\label{sec:summary}

In this paper we have discussed how to perform
Monte Carlo calculations
in quantum cosmology,
which enable us to go beyond the previous studies
based on the mini-superspace and saddle-point approximations.
In particular, we have adopted the recent proposal for defining quantum gravity
using the path integral over
the space-time metric with the Lorentzian signature,
where the oscillating integral is made well-defined
by the Picard-Lefschetz theory.
While the oscillating integral is difficult to deal with
by standard Monte Carlo methods
due to the sign problem,
we overcome this problem by applying the GTM, which deforms the integration contour
similarly to the Picard-Lefschetz theory.
As a first step, we have applied this method to the mini-superspace model.


As for the initial condition, we have followed
the no-boundary proposal, which corresponds
to requiring that the scale factor should vanish at the initial time.
This can be realized either by the Dirichlet
or the Robin boundary condition,
where in the latter case one has to
generalize the no-boundary proposal to
the vanishing \emph{expectation value} of the scale factor at the initial time.
In either case, the system undergoes quantum tunneling at early times, which
is represented by the purely imaginary expectation values of the
lapse function.\footnote{This is analogous
to ordinary quantum systems,
  where quantum tunneling is represented
  by complex trajectories in the real-time path integral \cite{Nishimura:2023dky}.}
The overall sign of the purely imaginary lapse function then
tells us
whether the relevant saddle point is of the Vilenkin type or
of the Hartle-Hawking type.
This is important since the different
orientation of the Wick rotation
may affect the stability of the tensor modes
and matter fields \cite{Feldbrugge:2017fcc,Feldbrugge:2017mbc}.

As expected from the semi-classical analysis, we find that
the relevant saddle points are
of the Vilenkin type for the Dirichlet boundary condition,
whereas they become the saddle points of the Hartle-Hawking type
for the
Robin boundary condition if the parameter $\tilde{\beta}$ is greater than
the critical value.
We pointed out, however, that
in the case of Robin boundary condition,
%
there is a problem in choosing the integration domain of the lapse function to be
the positive real axis since the expectation value of the lapse function tends to
vanish unless the scale factor at the final time is large enough.
While this can be avoided by extending
the integration domain to the whole real axis,
it remains to be seen whether this is an acceptable resolution to this issue.

After the quantum tunneling,
the real time emerges
as one can see from
the expectation value of the lapse function
that acquires a real part.
This happens due to the Stokes phenomenon, in which two saddle points
on the imaginary axis merge and split into two in the real direction.
Our simulation yields full quantum results, which are consistent with this picture.
The complex nature of the lapse function after the quantum tunneling
is simply due to the chosen gauge fixing for
the time-reparametrization invariance, in which
we require that the lapse function
be constant in time.
We discussed how to obtain
an equivalent saddle point configuration with
the lapse function
that is purely imaginary at early times
and
purely real at late times.

The saddle-point analysis of the previous work is justified when
the cosmological constant $\Lambda = 3 H^2$ is small.
We have evaluated the expectation value of the lapse function by the
$H^2$ expansion up to the fifth order and find that the perturbative regime is
around $H^2 \lesssim 0.1$.
This suggests that the perturbative expansion is not
valid any more with our choice $H^2=1$ for the simulation
although
the expectation value of the lapse function
and hence that of the scale factor turned out to be close to the semi-classical results.



With this capability to perform full quantum calculations,
we can investigate various problems in quantum cosmology
such as the instability issue of the Vilenkin saddle point
by adding the tensor modes and matter fields in the system,
which is straightforward.
It would be also interesting to perform simulations with different gauge fixing
for the time-reparametrization invariance so that one can obtain a real geometry
directly as the expectation value.
Last but not the least, we consider that quantum cosmology 
should eventually 
have a UV completion
in string theory. 
In particular, the simulation results of a nonperturbative formulation
of string theory, based on the type IIB matrix model \cite{Ishibashi:1996xs},
suggest the existence of 
the transition from Euclidean to Lorentzian geometry
at early times \cite{Nishimura:2022alt,Anagnostopoulos:2022dak,Hirasawa:2024dht}.
In this formulation,
there is no need to specify the initial condition
since even the time as well as the space emerges dynamically.
By comparing with the results obtained in this way,
it may be possible to determine the initial boundary condition
to be used in quantum cosmology.

\subsection*{Acknowledgements}

We would like to thank Masafumi Fukuma,
Hikaru Kawai, Jean-Luc Lehners and Kengo Shimada for valuable discussions.
The authors are also grateful to 
Katsuta Sakai and Atis Yosprakob
for their help concerning the technique developed in Ref.~\cite{Nishimura:2024bou}.
The computations were carried out on
the PC clusters in KEK Theory Center.
This work was supported by JST, the establishment of university fellowships towards the creation of science
technology innovation, Grant Number JPMJFS2136.
\appendix

\section{Brief review of the GTM}
\label{sec:GTM}

In this section, we briefly review the GTM \cite{Alexandru:2015sua},
which is used in our work in order to overcome the sign problem
that occurs in performing Monte Carlo simulations in quantum cosmology.
Let us consider a general model
\begin{alignat}{1}
Z=\int_{\bbR^n} dx\,e^{-S(x)}
\label{Z-general}
\end{alignat}
with $S\in\bbC$ and $x\in\bbR^n$
and define the expectation value of an observable $\mathcal{O}(x)$ as
\begin{alignat}{1}
\langle \mathcal{O}(x)\rangle={1\over Z}\int_{\bbR^n} dx\,e^{-S(x)}\mathcal{O}(x) \ .
\label{ex}
\end{alignat}
Since the phase of the complex weight $e^{-S(x)}$ oscillates as
a function of $x$, it is difficult to evaluate the expectation value
\eqref{ex} by ordinary Monte Carlo methods for a large system size $n$,
which is nothing but the sign problem.

The basic idea of the GTM is to
deform the integration contour
in such a way that the phase of the integrand does not
oscillate much even for large $n$.
This can be achieved by using the so-called anti-holomorphic gradient flow equation
\begin{alignat}{1}
    {\partial z_k(x,\tau) \over\partial\tau}
  &=\overline{\partial S(z(x,\tau))\over\partial z_k} \ ,
\label{floweq}
\end{alignat}
which defines
a one-to-one map from $z(x,0) \equiv x\in\bbR^n$ to $z(x,\tau)\in\bbC^n$.
Due to Cauchy's theorem, the integrals \eqref{Z-general} and (\ref{ex})
remain unaltered
under the deformation of the integration contour
from $\bbR^n$ to the $n$-dimensional
real manifold $M_\tau=\{z(x,\tau)\vert x\in\bbR^n\}$ embedded in $\bbC^n$,
which implies
\begin{alignat}{1}
  \langle \mathcal{O}(x)\rangle
  &={\int_{M_\tau} dz\,e^{-S(z)}\mathcal{O}(z)\over\int_{M_\tau} dz\,e^{-S(z)}} \ .
\end{alignat}
Note here that
$dz\,e^{-S(z)}$ can be decomposed as
\begin{alignat}{1}
dz\,e^{-S(z)}&=|dz|e^{i\phi(z)}\,e^{-\mathrm{Re}S(z)}e^{-i\mathrm{Im}S(z)}\nonumber\\
&=|dz|\,e^{-\mathrm{Re}S(z)}e^{i\theta(z)} \ ,
\label{dz-exp-S-decompose}
\end{alignat}
where $\phi(z)$ is the phase of $dz$ and
$e^{i\theta(z)}\equiv e^{i\phi(z)}e^{-i\mathrm{Im}S(z)}$.
Thus the expectation value can be evaluated by Monte Carlo simulation
using the reweighting formula as
\begin{alignat}{1}
  \langle \mathcal{O}(x)\rangle
  &= {\langle e^{i\theta(z)}\mathcal{O}(z)\rangle_{\tau}
   \over\langle e^{i\theta(z)}\rangle_{\tau}} \ ,
 \label{rew}
\end{alignat}
where the expectation value
$ \langle \ \cdot \ \rangle_{\tau}$ is defined with respect to
the partition function
\begin{alignat}{1}
  Z_{\tau}&=\int_{M_\tau}|dz|\,e^{-\mathrm{Re}S(z)} \ .
  \label{def-Z-tau}
\end{alignat}

The crucial property of the flow equation \eqref{floweq}
that
enables us to overcome the sign problem is that
\begin{alignat}{1}
  {\partial S(z(x,\tau))\over\partial \tau}
  &=
  {\partial S(z(x,\tau))\over\partial z_k}
  {\partial z_k(x,\tau) \over\partial\tau}
= \left|{\partial S(z(x,\tau))\over\partial z_k}\right|^2\ge 0 \ ,
  \label{flowinq}
\end{alignat}
which implies that the real part of the action increases
along the flow
while the imaginary part of the action remains constant.
Therefore, most of the points $x \in \bbR^n$ flow to some points
$z(x,\tau)\in\bbC^n$, for which $\mathrm{Re}S(z)$
is large,
and do not contribute to the integral \eqref{def-Z-tau}.
The only exceptions are the points $x \in \bbR^n$ that flow towards
some fix points defined by
${\partial S(z)\over\partial z_k}=0$
since $\mathrm{Re}S(z)$ will not increase much while the points are
flowing towards the fixed point.
In the $\tau\rightarrow\infty$ limit, in particular,
$M_\tau$ converges to a set of Lefschetz thimbles associated with
the fixed points
and
$\mathrm{Im}S(z)$ is constant on each thimble.\footnote{See
Ref.~\cite{Fujii:2013sra} for discussions on the residual
  sign problem coming from the integration measure
  represented by $\phi(z)$ in \eqref{dz-exp-S-decompose}.}
In practice, one can just make the flow time $\tau$
large enough to make the sign problem under control.


Extra care is needed if multiple thimbles exist
since the potential barrier between the regions associated with
each thimble
tends to diverge as the flow time $\tau$ increases,
which causes the ergodicity problem at large $\tau$.
In order to overcome this problem,
one can integrate over the flow time $\tau$
by treating it as one of the dynamical variables
in the simulation \cite{Fukuma:2020fez}.
This is not done in the present work
since we can perform meaningful calculations
by restricting ourselves to the region associated with one single thimble.
As we mentioned at the end of Section \ref{sec:discrete-model-robin}, however,
if one wishes to
sample configurations
from both
the $N>0$ and $N<0$ regions,
one needs to implement the integration over the flow time.



Monte Carlo simulation of the model \eqref{def-Z-tau} can be performed
by the HMC algorithm, which uses a fictitious Hamilton dynamics
to update the configuration
$z(x,\tau)$.
There are two different approaches
in defining the fictitious Hamilton dynamics.
One is to define a fictitious Hamilton dynamics of $z$ constrained on
the $n$-dimensional real manifold $M_\tau$ embedded
in $\bbC^n$ \cite{Fukuma:2019uot}.
The other is to define a fictitious Hamilton dynamics of
$x \in \bbR^n$ \cite{Fujisawa:2021hxh}.
Let us refer to these approaches as the on-thimble approach and the
on-axis approach, respectively, following Ref.~\cite{Nishimura:2024bou},
where pros and cons of each approach are discussed in detail.
Here we use the on-thimble approach, which has an advantage that
the modulus of the Jacobian for
the change of variables
associated with
the flow is included in the HMC procedure unlike the on-axes approach.
The only drawback of the on-thimble approach is that
one has to deal with a constrained Hamilton dynamics so that
the fictitious time-evolution is constrained to
the $n$-dimensional real manifold $M_\tau$
embedded in $\bbC^n$.

The procedure of the HMC algorithm 
on the deformed contour $M_\tau$
can be summarized as follows \cite{Fujii:2013sra,Fukuma:2019uot}.
\begin{enumerate}
   \item Starting from a configuration $z(x,\tau)\in M_\tau$, we generate $\tilde{p}\in\bbC^{n}$ with a Gaussian distribution $\propto \exp(-\tilde{p}^2/2)$.
   \item We project $\tilde{p}$ onto the tangent space of $M_\tau$ at $z(x,\tau)$ as
       \begin{alignat}{1}
       p&=J(x,\tau)\, \mathrm{Re}(J^{-1}(x,\tau)\tilde{p})
       \end{alignat}
     with the Jacobian 
       \begin{alignat}{1}
       J_{kl}(x,\tau)&={\partial z_k(x,\tau)\over \partial x_l} \ .
       \end{alignat}
     \item 
       With the $z$ and $p$ given above,
       we search for $u,\,\lambda\in\bbR^n$ satisfying
       \begin{alignat}{1}
         z_k(x+u,\tau)-z_k(x,\tau)
         -p_k\Delta s
         +{\Delta s^2\over2}{\partial \mathrm{Re}S(z(x,\tau))\over\partial z_k}
         +iJ_{kl}(z(x,\tau))\lambda_l=0 \ ,
       \end{alignat}
       where $\Delta s$ is the step size. Then set $x'=x+u$.
   \item Define 
       \begin{alignat}{1}
         \tilde{p}'_k
         ={1\over\Delta s}(z_k(x',\tau)-z_k(x,\tau))
         -{\Delta s\over2}{ \partial \mathrm{Re}S(z(x',\tau)) \over \partial z_k}
       \end{alignat}
       and project $\tilde{p}'$ onto the tangent space of $M_\tau$
       at $z(x',\tau)$ as
       \begin{alignat}{1}
       p'=J(x',\tau) \, \mathrm{Re}(J^{-1}(x',\tau)\tilde{p}') \ .
       \end{alignat}
     \item We repeat the steps 3 to 4 for a fixed number of times, which
       we denote as $n_{\rm s}$.
     \item
      We update the configuration from $z(x,\tau)$ to $z(x',\tau)$
      with the probability $\mathrm{min}\left(1,e^{-\Delta H}\right)$,
      where $\Delta H$ is the change of Hamiltonian defined as
       \begin{alignat}{2}
       \Delta H&=H(z(x',\tau),p')-H(z(x,\tau),p) \ , \quad
       \mbox{where~} & H(z,p)= \frac{1}{2} \, p^2 + \mathrm{Re}S(z) \ .
       \end{alignat}
       This is the usual Metropolis test in the HMC algorithm, which is needed to
       guarantee the detailed balance.
\end{enumerate}

       The two parameters $\Delta s$ and $n_{\rm s}$ in the HMC algorithm
       can be optimized in the standard manner.
       First we optimize the step size $\Delta s$
       for fixed $s_{\rm f}\equiv n_{\rm s} \Delta s$
       by maximizing the product
       $\Delta s \cdot P_{\rm acc}(\Delta s)$, where $P_{\rm acc}(\Delta s)$ is
       the acceptance rate of the Metropolis test.
       Then the total time $s_{\rm f}$
       of the fictitious Hamilton dynamics is optimized
       by minimizing the computational time 
       needed to obtain a
       decorrelated configuration.
In Table \ref{tab:parameters},
we show the parameters used in our simulations.

\begin{table}[t]
\centering
\begin{tabular}{c|c|c|c|c|c|c|c}
\hline
b.c.\      & $H^2$ & $q_f$   & $\tilde{\beta}/\tilde{\beta}_c$    & $\tau$    & $n_\tau$   & $s_{\rm f}$ &
$n_s$   \\ \hline
D &  1 & 0.8 & ---   & 4   & 40 & 0.1 &  15 \\
D &  1 & 10 & ---  & 4  & 40 & 0.38 & 10 \\
R &  1 & 10 & 0.3   & 4    & 40   & 0.39 & 10    \\
R &  1 & 10  & 1.2 & 6 & 60 & 0.4 & 10 \\
R &  0.3 & 10  & 1.2 & 5 & 50 & 0.25 & 10 \\
R &  0.1 & 10  & 1.2 & 5 & 50 & 0.25& 20 \\
R &  0.03& 10  & 1.2 & 5 & 70 & 0.2 & 30 \\
\hline
\end{tabular}
\caption{The parameters for the HMC simulation by the GTM are presented.
  The D and R
  in the first column represent the
  Dirichlet and Robin boundary conditions, respectively, at the initial time.
  In the fifth and sixth columns, we show the parameters used in solving
  the flow equation, where $\tau$  represents the total flow time and $n_\tau$ represents
  the number of steps.
  In the last two
  columns,
  we show the parameters used in the
  HMC algorithm, where
  $s_{\rm f}=  n_{\rm s} \Delta s$
  represents the total time of the fictitious Hamilton dynamics
  and $n_{\rm s}$ represents the number of steps used in solving the Hamilton equation.}
\label{tab:parameters}
\end{table}

Simulation results presented in Section \ref{sec:results}
have been obtained by
solving the fictitious Hamilton dynamics 100,000 times
and saving the configuration every ten times.
As
in any Monte Carlo methods,
once we generate a set of configurations $\{z^{(j)}\}_{j=1,\ldots,N_{\mathrm{conf}}}$,
we can estimate the expectation value as
\begin{alignat}{2}
  \langle f(z)\rangle_{\tau}\approx{1\over N_{\mathrm{conf}}}\sum_{j=1}^{N_{\mathrm{conf}}}
  f(z^{(j)})\equiv\overline{f(z)} \ . \label{approx}
\end{alignat}
Plugging (\ref{approx}) into (\ref{rew}), one finds
\begin{alignat}{2}
\langle \mathcal{O}(z)\rangle\approx{\overline{e^{i\theta(z)}\mathcal{O}(z)}\over\overline{e^{i\theta(z)}}} \ . \label{exp-avg}
\end{alignat}
Thus one can obtain
the expectation value through (\ref{exp-avg}) up to some statistical error, which
is of the order of O($1/\sqrt{N_{\mathrm{conf}}}$)
if the sampled configurations are separated well enough
compared with the autocorrelation time.
We estimate the statistical error
using the jackknife method, which takes into account the correlation
among the sampled configurations.

Next we discuss a technical problem
with the flow equation \eqref{floweq}, which was pointed out recently
in Ref.~\cite{Nishimura:2024bou}.
For that,
let us define the Hessian of the action as
\begin{alignat}{2}
 H_{kl}={\partial^2 S(z)\over \partial z_k\partial z_l} \ ,
\end{alignat}
and consider the singular value decomposition of the symmetric matrix $H$,
\begin{alignat}{2}
  H &= U^T D  \, U  \ ,
  \label{svd}
\end{alignat}
where $U$ is a unitary matrix and $D={\rm diag}(d_1,\ldots, d_n)$
is a real positive diagonal matrix.
Then, the stiffness of the system can be defined as
$\eta(H)={d_n\over d_1}$.
High stiffness makes the effect of the flow equation very different
for each mode,
which causes difficulty in the 
simulation.\footnote{This problem occurs, in particular,
  when we choose the parameter $\tilde{\beta}$
  in the Robin boundary condition at the initial time
  to be $\tilde{\beta} > \tilde{\beta}_{\rm c}$.}

In order to solve this problem,
we have used a new technique \cite{Nishimura:2024bou},
which introduces a ``preconditioner'' in the flow equation as
\begin{alignat}{1}
 {\partial\over\partial\tau}z_k(x,\tau)
= A_{kl}\overline{\partial S(z(x,\tau))\over\partial z_l}
 \ ,
\label{prefloweq}
\end{alignat}
where $A$ should be a positive-definite hermitian matrix
in order to keep the crucial property (\ref{flowinq}) of the flow equation intact.
Making use of this freedom, 
we can optimize the stiffness ($\eta(\bar{A} H)=1$)
by choosing 
\begin{alignat}{1}
A(z,\bar{z})=U^\dag  D^{-1} U = (\bar{H}(\bar{z})H(z))^{-{1\over2}} \ .
\end{alignat}
When we solve the flow equation numerically, we discretize the flow equation
using the so-called ``Runge-Kutta 4'' method.


\section{Derivation of the effective action at finite $\epsilon$}
\label{sec:eff-action-epsilon}

In this section, we derive the effective action \eqref{action:eff-r}
for the lapse function $N$
at finite $\epsilon$, which is used in obtaining the
perturbative expansion \eqref{Nav-sim}.
It is useful to consider first the case of Dirichlet boundary condition
with $q(0)=q_{\rm i}$ left arbitrary.

Let us start from the discretized partition function \eqref{discretized-Z-DD}.
Solving the classical equation of motion \eqref{qEOM-discrete}
with the boundary condition $q_0 = q_{\rm i}$ and $q_n = q_{\rm f}$,
we get
\begin{align}
  q_k &= q_{\rm i} + k(k+1) \epsilon^2 N^2  + C \, \frac{k}{n}  \ ,
  \quad\quad
  \mbox{where~}
  C = q_{\rm f} - q_{\rm i} - n(n+1) \epsilon^2 N^2 \ .
\end{align}
We can integrate out $q_k$ ($k=1 , \cdots , (n-1) $)
by just performing the Gaussian integral,
which amounts to plugging the classical solution in the action
and getting the factors $(\sqrt{N})^{n-1}$.
Thus
we obtain
\begin{alignat}{2}
Z &= \int dN \, N^{-1/2} \, e^{iS_{\rm eff}^{\rm (D)}/H^2} \ ,
\\
S_{\rm eff}^{\rm (D)} &=
- \frac{\pi ^2 }{2}\left\{
\epsilon^2 N^3  - N^3 + 6 N (q_{\rm f} + q_{\rm i} - 2 )
+ \frac{3}{N} (q_{\rm f} - q_{\rm i})^2 \right\} \ ,
\label{action:eff-D-discrete}
\end{alignat}
up to an irrelevant constant.
The finite $\epsilon$ effect appears only as 
an ${\rm O}(\epsilon^2)$ term in the effective
action \eqref{action:eff-D-discrete},
and we retrieve
\eqref{eq:eff-S-DD}
in the $\epsilon \rightarrow 0$ limit.

In the case of Robin boundary condition, we 
start from the discretized partition function \eqref{action-robin-q}.
The integration over $q_k$ except for $q_0$
can be readily done as in the Dirichlet case. Thus we are left with
the integration over $q_0= q_{\rm i}$ given by
\begin{alignat}{2}
  Z &=\int dN \, N^{-1/2}
  \int dq_{\rm i}  \, e^{i \{ S_{\rm eff}^{\rm (D)}
    +  (\alpha q_{\rm i} + \frac{1}{2\beta} (q_{\rm i})^2 \} /H^2 } \ .
\label{action:eff-D-discrete-Robin}
\end{alignat}
Extracting the terms that depend on $q_{\rm i}$, we get
\begin{align}
  S_{\rm eff}
  &=
  \frac{N-3 \pi^2 \beta}{2 \beta N} (q_{\rm i})^2  
    + \left\{ \frac{3 \pi^2}{N}  (q_{\rm f} - N^2) + \alpha \right\}
    q_{\rm i}
    \ .
\end{align}
The Gaussian integral over $q_{\rm i}$ yields an extra factor
$\sqrt{N/(N-3 \pi^2 \beta)}$.
Thus we arrive at
\begin{alignat}{2}
  Z &= \int dN \, (N-3 \pi^2 \beta)^{-1/2}
  e^{iS_{\rm eff}^{\rm (R)}/H^2} \ , \\
  S_{\rm eff}^{\rm (R)} &=
-  \frac{\pi^2}{2}\left[ \epsilon ^2 N^3 + 
    \frac{\beta N\left\{ \frac{3 \pi^2}{N} (q_{\rm f}-N^2) + \alpha\right\}^2 }
         {\pi^2 (N- 3\pi^2 \beta)}
         - \left\{ N^3 - 6 N (q_{\rm f}-2) - \frac{3}{N} (q_{\rm f})^2  \right\}
\right] \ ,
    \label{action:eff-discrete-Robin}
\end{alignat}
which leads to \eqref{action:eff-r}.
In the $\epsilon \rightarrow 0$ limit, we retrieve the result
\eqref{action-eff-r} in the continuum.


\bibliographystyle{JHEP}
\bibliography{thimble-qc}

\providecommand{\href}[2]{#2}\begingroup\raggedright\begin{thebibliography}{10}

\bibitem{Vilenkin:1982de}
A.~Vilenkin, \emph{{Creation of Universes from Nothing}}, \href{https://doi.org/10.1016/0370-2693(82)90866-8}{\emph{Phys. Lett. B} {\bfseries 117} (1982) 25}.

\bibitem{Vilenkin:1984wp}
A.~Vilenkin, \emph{{Quantum Creation of Universes}}, \href{https://doi.org/10.1103/PhysRevD.30.509}{\emph{Phys. Rev. D} {\bfseries 30} (1984) 509}.

\bibitem{Vilenkin:1994rn}
A.~Vilenkin, \emph{{Approaches to quantum cosmology}}, \href{https://doi.org/10.1103/PhysRevD.50.2581}{\emph{Phys. Rev. D} {\bfseries 50} (1994) 2581} [\href{https://arxiv.org/abs/gr-qc/9403010}{{\ttfamily gr-qc/9403010}}].

\bibitem{Hartle:1983ai}
J.B.~Hartle and S.W.~Hawking, \emph{{Wave Function of the Universe}}, \href{https://doi.org/10.1103/PhysRevD.28.2960}{\emph{Phys. Rev. D} {\bfseries 28} (1983) 2960}.

\bibitem{Lehners:2023yrj}
J.-L.~Lehners, \emph{{Review of the no-boundary wave function}}, \href{https://doi.org/10.1016/j.physrep.2023.06.002}{\emph{Phys. Rept.} {\bfseries 1022} (2023) 1} [\href{https://arxiv.org/abs/2303.08802}{{\ttfamily 2303.08802}}].

\bibitem{Maldacena:2024uhs}
J.~Maldacena, \emph{{Comments on the no boundary wavefunction and slow roll inflation}},  \href{https://arxiv.org/abs/2403.10510}{{\ttfamily 2403.10510}}.

\bibitem{Halliwell:1988ik}
J.J.~Halliwell and J.~Louko, \emph{{Steepest Descent Contours in the Path Integral Approach to Quantum Cosmology. 1. The De Sitter Minisuperspace Model}}, \href{https://doi.org/10.1103/PhysRevD.39.2206}{\emph{Phys. Rev. D} {\bfseries 39} (1989) 2206}.

\bibitem{Gibbons:1978ac}
G.W.~Gibbons, S.W.~Hawking and M.J.~Perry, \emph{{Path Integrals and the Indefiniteness of the Gravitational Action}}, \href{https://doi.org/10.1016/0550-3213(78)90161-X}{\emph{Nucl. Phys. B} {\bfseries 138} (1978) 141}.

\bibitem{Feldbrugge:2017kzv}
J.~Feldbrugge, J.-L.~Lehners and N.~Turok, \emph{{Lorentzian Quantum Cosmology}}, \href{https://doi.org/10.1103/PhysRevD.95.103508}{\emph{Phys. Rev. D} {\bfseries 95} (2017) 103508} [\href{https://arxiv.org/abs/1703.02076}{{\ttfamily 1703.02076}}].

\bibitem{DiazDorronsoro:2017hti}
J.~Diaz~Dorronsoro, J.J.~Halliwell, J.B.~Hartle, T.~Hertog and O.~Janssen, \emph{{Real no-boundary wave function in Lorentzian quantum cosmology}}, \href{https://doi.org/10.1103/PhysRevD.96.043505}{\emph{Phys. Rev. D} {\bfseries 96} (2017) 043505} [\href{https://arxiv.org/abs/1705.05340}{{\ttfamily 1705.05340}}].

\bibitem{DiazDorronsoro:2018wro}
J.~Diaz~Dorronsoro, J.J.~Halliwell, J.B.~Hartle, T.~Hertog, O.~Janssen and Y.~Vreys, \emph{{Damped perturbations in the no-boundary state}}, \href{https://doi.org/10.1103/PhysRevLett.121.081302}{\emph{Phys. Rev. Lett.} {\bfseries 121} (2018) 081302} [\href{https://arxiv.org/abs/1804.01102}{{\ttfamily 1804.01102}}].

\bibitem{Halliwell:2018ejl}
J.J.~Halliwell, J.B.~Hartle and T.~Hertog, \emph{{What is the No-Boundary Wave Function of the Universe?}}, \href{https://doi.org/10.1103/PhysRevD.99.043526}{\emph{Phys. Rev. D} {\bfseries 99} (2019) 043526} [\href{https://arxiv.org/abs/1812.01760}{{\ttfamily 1812.01760}}].

\bibitem{Chen:2023prz}
H.-Y.~Chen, Y.~Hikida, Y.~Taki and T.~Uetoko, \emph{{Complex saddles of three-dimensional de Sitter gravity via holography}}, \href{https://doi.org/10.1103/PhysRevD.107.L101902}{\emph{Phys. Rev. D} {\bfseries 107} (2023) L101902} [\href{https://arxiv.org/abs/2302.09219}{{\ttfamily 2302.09219}}].

\bibitem{Chen:2023sry}
H.-Y.~Chen, Y.~Hikida, Y.~Taki and T.~Uetoko, \emph{{Complex saddles of Chern-Simons gravity and dS3/CFT2 correspondence}}, \href{https://doi.org/10.1103/PhysRevD.108.066005}{\emph{Phys. Rev. D} {\bfseries 108} (2023) 066005} [\href{https://arxiv.org/abs/2306.03330}{{\ttfamily 2306.03330}}].

\bibitem{Chen:2024vpa}
H.-Y.~Chen, Y.~Hikida, Y.~Taki and T.~Uetoko, \emph{{Semi-classical saddles of three-dimensional gravity via holography}},  \href{https://arxiv.org/abs/2403.02108}{{\ttfamily 2403.02108}}.

\bibitem{Chen:2024qmn}
H.-Y.~Chen, Y.~Hikida, Y.~Taki and T.~Uetoko, \emph{{The semi-classical saddles in three-dimensional gravity via holography and mini-superspace approach}},  \href{https://arxiv.org/abs/2404.10277}{{\ttfamily 2404.10277}}.

\bibitem{Feldbrugge:2017fcc}
J.~Feldbrugge, J.-L.~Lehners and N.~Turok, \emph{{No smooth beginning for spacetime}}, \href{https://doi.org/10.1103/PhysRevLett.119.171301}{\emph{Phys. Rev. Lett.} {\bfseries 119} (2017) 171301} [\href{https://arxiv.org/abs/1705.00192}{{\ttfamily 1705.00192}}].

\bibitem{Feldbrugge:2017mbc}
J.~Feldbrugge, J.-L.~Lehners and N.~Turok, \emph{{No rescue for the no boundary proposal: Pointers to the future of quantum cosmology}}, \href{https://doi.org/10.1103/PhysRevD.97.023509}{\emph{Phys. Rev. D} {\bfseries 97} (2018) 023509} [\href{https://arxiv.org/abs/1708.05104}{{\ttfamily 1708.05104}}].

\bibitem{Vilenkin:2018dch}
A.~Vilenkin and M.~Yamada, \emph{{Tunneling wave function of the universe}}, \href{https://doi.org/10.1103/PhysRevD.98.066003}{\emph{Phys. Rev. D} {\bfseries 98} (2018) 066003} [\href{https://arxiv.org/abs/1808.02032}{{\ttfamily 1808.02032}}].

\bibitem{Vilenkin:2018oja}
A.~Vilenkin and M.~Yamada, \emph{{Tunneling wave function of the universe II: the backreaction problem}}, \href{https://doi.org/10.1103/PhysRevD.99.066010}{\emph{Phys. Rev. D} {\bfseries 99} (2019) 066010} [\href{https://arxiv.org/abs/1812.08084}{{\ttfamily 1812.08084}}].

\bibitem{Feldbrugge:2018gin}
J.~Feldbrugge, J.-L.~Lehners and N.~Turok, \emph{{Inconsistencies of the New No-Boundary Proposal}}, \href{https://doi.org/10.3390/universe4100100}{\emph{Universe} {\bfseries 4} (2018) 100} [\href{https://arxiv.org/abs/1805.01609}{{\ttfamily 1805.01609}}].

\bibitem{Matsui:2022lfj}
H.~Matsui, S.~Mukohyama and A.~Naruko, \emph{{No smooth spacetime in Lorentzian quantum cosmology and trans-Planckian physics}}, \href{https://doi.org/10.1103/PhysRevD.107.043511}{\emph{Phys. Rev. D} {\bfseries 107} (2023) 043511} [\href{https://arxiv.org/abs/2211.05306}{{\ttfamily 2211.05306}}].

\bibitem{Matsui:2023hei}
H.~Matsui and S.~Mukohyama, \emph{{Hartle-Hawking no-boundary proposal and Ho\v{r}ava-Lifshitz gravity}}, \href{https://doi.org/10.1103/PhysRevD.109.023504}{\emph{Phys. Rev. D} {\bfseries 109} (2024) 023504} [\href{https://arxiv.org/abs/2310.00210}{{\ttfamily 2310.00210}}].

\bibitem{Matsui:2024bfn}
H.~Matsui, \emph{{No smooth spacetime: Exploring primordial perturbations in Lorentzian quantum cosmology}},  \href{https://arxiv.org/abs/2404.18609}{{\ttfamily 2404.18609}}.

\bibitem{DiTucci:2019dji}
A.~Di~Tucci and J.-L.~Lehners, \emph{{No-Boundary Proposal as a Path Integral with Robin Boundary Conditions}}, \href{https://doi.org/10.1103/PhysRevLett.122.201302}{\emph{Phys. Rev. Lett.} {\bfseries 122} (2019) 201302} [\href{https://arxiv.org/abs/1903.06757}{{\ttfamily 1903.06757}}].

\bibitem{DiTucci:2019bui}
A.~Di~Tucci, J.-L.~Lehners and L.~Sberna, \emph{{No-boundary prescriptions in Lorentzian quantum cosmology}}, \href{https://doi.org/10.1103/PhysRevD.100.123543}{\emph{Phys. Rev. D} {\bfseries 100} (2019) 123543} [\href{https://arxiv.org/abs/1911.06701}{{\ttfamily 1911.06701}}].

\bibitem{Alexandru:2015sua}
A.~Alexandru, G.~Basar, P.F.~Bedaque, G.W.~Ridgway and N.C.~Warrington, \emph{{Sign problem and Monte Carlo calculations beyond Lefschetz thimbles}}, \href{https://doi.org/10.1007/JHEP05(2016)053}{\emph{JHEP} {\bfseries 05} (2016) 053} [\href{https://arxiv.org/abs/1512.08764}{{\ttfamily 1512.08764}}].

\bibitem{Witten:2010cx}
E.~Witten, \emph{{Analytic continuation of Chern-Simons theory}}, {\emph{AMS/IP Stud. Adv. Math.} {\bfseries 50} (2011) 347} [\href{https://arxiv.org/abs/1001.2933}{{\ttfamily 1001.2933}}].

\bibitem{Cristoforetti:2012su}
{\scshape AuroraScience} collaboration, \emph{{New approach to the sign problem in quantum field theories: High density QCD on a Lefschetz thimble}}, \href{https://doi.org/10.1103/PhysRevD.86.074506}{\emph{Phys. Rev. D} {\bfseries 86} (2012) 074506} [\href{https://arxiv.org/abs/1205.3996}{{\ttfamily 1205.3996}}].

\bibitem{Cristoforetti:2013wha}
M.~Cristoforetti, F.~Di~Renzo, A.~Mukherjee and L.~Scorzato, \emph{{Monte Carlo simulations on the Lefschetz thimble: Taming the sign problem}}, \href{https://doi.org/10.1103/PhysRevD.88.051501}{\emph{Phys. Rev. D} {\bfseries 88} (2013) 051501} [\href{https://arxiv.org/abs/1303.7204}{{\ttfamily 1303.7204}}].

\bibitem{Fujii:2013sra}
H.~Fujii, D.~Honda, M.~Kato, Y.~Kikukawa, S.~Komatsu and T.~Sano, \emph{{Hybrid Monte Carlo on Lefschetz thimbles - A study of the residual sign problem}}, \href{https://doi.org/10.1007/JHEP10(2013)147}{\emph{JHEP} {\bfseries 10} (2013) 147} [\href{https://arxiv.org/abs/1309.4371}{{\ttfamily 1309.4371}}].

\bibitem{Loll:2019rdj}
R.~Loll, \emph{{Quantum Gravity from Causal Dynamical Triangulations: A Review}}, \href{https://doi.org/10.1088/1361-6382/ab57c7}{\emph{Class. Quant. Grav.} {\bfseries 37} (2020) 013002} [\href{https://arxiv.org/abs/1905.08669}{{\ttfamily 1905.08669}}].

\bibitem{Jia:2021xeh}
D.~Jia, \emph{{Complex, Lorentzian, and Euclidean simplicial quantum gravity: numerical methods and physical prospects}}, \href{https://doi.org/10.1088/1361-6382/ac4b04}{\emph{Class. Quant. Grav.} {\bfseries 39} (2022) 065002} [\href{https://arxiv.org/abs/2110.05953}{{\ttfamily 2110.05953}}].

\bibitem{Jia:2022nda}
D.~Jia, \emph{{Truly Lorentzian quantum cosmology}}, \href{https://doi.org/10.1103/PhysRevD.108.103540}{\emph{Phys. Rev. D} {\bfseries 108} (2023) 103540} [\href{https://arxiv.org/abs/2211.00517}{{\ttfamily 2211.00517}}].

\bibitem{Fukuma:2019uot}
M.~Fukuma, N.~Matsumoto and N.~Umeda, \emph{{Implementation of the HMC algorithm on the tempered Lefschetz thimble method}},  \href{https://arxiv.org/abs/1912.13303}{{\ttfamily 1912.13303}}.

\bibitem{Nishimura:2024bou}
J.~Nishimura, K.~Sakai and A.~Yosprakob, \emph{{Preconditioned flow as a solution to the hierarchical growth problem in the generalized Lefschetz thimble method}}, \href{https://doi.org/10.1007/JHEP07(2024)174}{\emph{JHEP} {\bfseries 07} (2024) 174} [\href{https://arxiv.org/abs/2404.16589}{{\ttfamily 2404.16589}}].

\bibitem{Nishimura:2023dky}
J.~Nishimura, K.~Sakai and A.~Yosprakob, \emph{{A new picture of quantum tunneling in the real-time path integral from Lefschetz thimble calculations}}, \href{https://doi.org/10.1007/JHEP09(2023)110}{\emph{JHEP} {\bfseries 09} (2023) 110} [\href{https://arxiv.org/abs/2307.11199}{{\ttfamily 2307.11199}}].

\bibitem{Teitelboim:1981ua}
C.~Teitelboim, \emph{{Quantum Mechanics of the Gravitational Field}}, \href{https://doi.org/10.1103/PhysRevD.25.3159}{\emph{Phys. Rev. D} {\bfseries 25} (1982) 3159}.

\bibitem{Teitelboim:1983fh}
C.~Teitelboim, \emph{{Causality Versus Gauge Invariance in Quantum Gravity and Supergravity}}, \href{https://doi.org/10.1103/PhysRevLett.50.705}{\emph{Phys. Rev. Lett.} {\bfseries 50} (1983) 705}.

\bibitem{Teitelboim:1983fk}
C.~Teitelboim, \emph{{The Proper Time Gauge in Quantum Theory of Gravitation}}, \href{https://doi.org/10.1103/PhysRevD.28.297}{\emph{Phys. Rev. D} {\bfseries 28} (1983) 297}.

\bibitem{Banihashemi:2024aal}
B.~Banihashemi and T.~Jacobson, \emph{{On the lapse contour}},  \href{https://arxiv.org/abs/2405.10307}{{\ttfamily 2405.10307}}.

\bibitem{Honda:2024aro}
M.~Honda, H.~Matsui, K.~Okabayashi and T.~Terada, \emph{{Resurgence in Lorentzian quantum cosmology: no-boundary saddles and resummation of quantum gravity corrections around tunneling saddles}},  \href{https://arxiv.org/abs/2402.09981}{{\ttfamily 2402.09981}}.

\bibitem{Ailiga:2023wzl}
M.~Ailiga, S.~Mallik and G.~Narain, \emph{{Lorentzian Robin Universe}}, \href{https://doi.org/10.1007/JHEP01(2024)124}{\emph{JHEP} {\bfseries 01} (2024) 124} [\href{https://arxiv.org/abs/2308.01310}{{\ttfamily 2308.01310}}].

\bibitem{Ailiga:2024mmt}
M.~Ailiga, S.~Mallik and G.~Narain, \emph{{Lorentzian path-integral of Robin Universe}},  \href{https://arxiv.org/abs/2407.16692}{{\ttfamily 2407.16692}}.

\bibitem{Fukuma:2020fez}
M.~Fukuma and N.~Matsumoto, \emph{{Worldvolume approach to the tempered Lefschetz thimble method}}, \href{https://doi.org/10.1093/ptep/ptab010}{\emph{PTEP} {\bfseries 2021} (2021) 023B08} [\href{https://arxiv.org/abs/2012.08468}{{\ttfamily 2012.08468}}].

\bibitem{Fujisawa:2021hxh}
G.~Fujisawa, J.~Nishimura, K.~Sakai and A.~Yosprakob, \emph{{Backpropagating Hybrid Monte Carlo algorithm for fast Lefschetz thimble calculations}}, \href{https://doi.org/10.1007/JHEP04(2022)179}{\emph{JHEP} {\bfseries 04} (2022) 179} [\href{https://arxiv.org/abs/2112.10519}{{\ttfamily 2112.10519}}].

\bibitem{Lehners:2021jmv}
J.-L.~Lehners, \emph{{Wave function of simple universes analytically continued from negative to positive potentials}}, \href{https://doi.org/10.1103/PhysRevD.104.063527}{\emph{Phys. Rev. D} {\bfseries 104} (2021) 063527} [\href{https://arxiv.org/abs/2105.12075}{{\ttfamily 2105.12075}}].

\bibitem{Witten:2021nzp}
E.~Witten, \emph{{A Note On Complex Spacetime Metrics}},  \href{https://arxiv.org/abs/2111.06514}{{\ttfamily 2111.06514}}.

\bibitem{Jonas:2022uqb}
C.~Jonas, J.-L.~Lehners and J.~Quintin, \emph{{Uses of complex metrics in cosmology}}, \href{https://doi.org/10.1007/JHEP08(2022)284}{\emph{JHEP} {\bfseries 08} (2022) 284} [\href{https://arxiv.org/abs/2205.15332}{{\ttfamily 2205.15332}}].

\bibitem{Hertog:2023vot}
T.~Hertog, O.~Janssen and J.~Karlsson, \emph{{Kontsevich-Segal Criterion in the No-Boundary State Constrains Inflation}}, \href{https://doi.org/10.1103/PhysRevLett.131.191501}{\emph{Phys. Rev. Lett.} {\bfseries 131} (2023) 191501} [\href{https://arxiv.org/abs/2305.15440}{{\ttfamily 2305.15440}}].

\bibitem{Ishibashi:1996xs}
N.~Ishibashi, H.~Kawai, Y.~Kitazawa and A.~Tsuchiya, \emph{{A Large N reduced model as superstring}}, \href{https://doi.org/10.1016/S0550-3213(97)00290-3}{\emph{Nucl. Phys. B} {\bfseries 498} (1997) 467} [\href{https://arxiv.org/abs/hep-th/9612115}{{\ttfamily hep-th/9612115}}].

\bibitem{Nishimura:2022alt}
J.~Nishimura, \emph{{Signature change of the emergent space-time in the IKKT matrix model}}, \href{https://doi.org/10.22323/1.406.0255}{\emph{PoS} {\bfseries CORFU2021} (2022) 255} [\href{https://arxiv.org/abs/2205.04726}{{\ttfamily 2205.04726}}].

\bibitem{Anagnostopoulos:2022dak}
K.N.~Anagnostopoulos, T.~Azuma, K.~Hatakeyama, M.~Hirasawa, Y.~Ito, J.~Nishimura et~al., \emph{{Progress in the numerical studies of the type IIB matrix model}}, \href{https://doi.org/10.1140/epjs/s11734-023-00849-x}{\emph{Eur. Phys. J. ST} {\bfseries 232} (2023) 3681} [\href{https://arxiv.org/abs/2210.17537}{{\ttfamily 2210.17537}}].

\bibitem{Hirasawa:2024dht}
M.~Hirasawa, K.N.~Anagnostopoulos, T.~Azuma, K.~Hatakeyama, J.~Nishimura, S.~Papadoudis et~al., \emph{{The effects of SUSY on the emergent spacetime in the Lorentzian type IIB matrix model}}, \href{https://doi.org/10.22323/1.463.0257}{\emph{PoS} {\bfseries CORFU2023} (2024) 257} [\href{https://arxiv.org/abs/2407.03491}{{\ttfamily 2407.03491}}].

\end{thebibliography}\endgroup


\end{document}